\documentclass[a4paper]{article}

\usepackage{upgreek}
\usepackage{amsmath}
\usepackage[ansinew]{inputenc}
\usepackage{makeidx}
\usepackage{wasysym}
\usepackage{natbib}
\usepackage{textcomp}
\usepackage{rotating}
\usepackage{amssymb}
\usepackage{url}
\usepackage{exscale}
\usepackage{icomma}
\usepackage{ifthen}
\usepackage{booktabs}
\usepackage{xspace}
\usepackage{color}
\def\comment#1{}

\newcommand{\m}[1]{\mathrm{#1}}

\newcommand{\V}[1]{\boldsymbol{\m{#1}}}

\newcommand{\vbeta}{\V \beta}
\newcommand{\vxi}{\V \xi}

\def\bn{\bigskip\noindent}

\def\va{{\bf a}}
\def\vb{{\bf b}}
\def\vd{{\bf d}}
\def\ve{{\bf e}}

\def\vg{{\bf g}}

\def\vI{{\bf I}}

\def\vk{{\bf k}}
\def\vL{{\bf L}}

\def\vn{{\bf n}}

\def\vr{{\bf r}}

\def\vv{{\bf v}}
\def\vw{{\bf w}}
\def\vW{{\bf W}}
\def\vx{{\bf x}}
\def\vz{{\bf z}}

\def\vA{{\bf A}}
\def\vB{{\bf B}}

\def\vF{{\bf F}}

\def\vR{{\bf R}}
\def\vS{{\bf S}}
\def\vV{{\bf V}}
\def\vX{{\bf X}}

\def\cE{{\cal E}}

\def\cO{{\cal O}}

\def\cT{{\cal T}}

\def\rA{{\rm A}}

\def\rE{{\rm E}}

\def\rN{{\rm N}}

\newfont{\mathe}{cmbx10}

\def\3{{\ss }}

\def\gbar{{\overline g}{}}

\def\wbar{{\overline w}{}}

\def\nhat{{\hat n}}

\def\la{\label}
\def\be{\begin{equation}}
\def\ee{\end{equation}}
\def\ba{\begin{eqnarray}}
\def\ea{\end{eqnarray}}

\newcommand{\block}{\Box}

\def\rL{{\rm L}}

\begin{document}

\ \vskip1truecm

\centerline{\huge\bf Advanced relativistic VLBI model} \bigskip\centerline{\huge\bf for geodesy}  \vskip0.5truecm
\centerline{Michael Soffel$^1$, Sergei Kopeikin$^{2,3}$ and Wen-Biao Han$^4$} \vskip0.5truecm

\noindent
{1. Lohrmann Observatory, Dresden Technical University, 01062 Dresden, Germany}

\noindent
{2. Department of Physics and Astronomy, University of Missouri,  Columbia, MO, 65211, USA}

\noindent
{3. Siberian State University of Geosystems and Technologies, Plakhotny Street 10, Novosibirsk 630108, Russia}

\noindent
{4. Shanghai Astronomical Observatory, Chinese Academy of Sciences, Shanghai, 200030, China}

\thispagestyle{empty}

\bigskip\noindent

{\bf Abstract:}

\medskip\noindent
Our present relativistic part of the geodetic VLBI model for Earthbound antennas is a consensus model which is considered as a standard for processing high-precision VLBI observations. It was created as a compromise between a variety of relativistic VLBI models proposed by different authors as documented in the IERS Conventions 2010. The accuracy of the consensus model is in the picosecond range for the group delay but this is not sufficient for current geodetic pur- poses. This paper provides a fully documented derivation of a new relativistic model having an accuracy substantially higher than one picosecond and based upon a well accepted formalism of relativistic celestial mechanics, astrometry and geodesy. Our new model fully confirms the consensus model at the picosecond level and in several respects goes to a great extent beyond it. More specifically, terms related to the acceleration of the geocenter are considered and kept in the model, the gravitational time-delay due to a massive body (planet, Sun, etc.) with arbitrary mass and spin-multipole moments is derived taking into account the motion of the body, and a new formalism for the time-delay problem of radio sources located at finite distance from VLBI stations is presented. Thus, the paper presents a substantially elaborated theoretical justification of the consensus model and its significant extension that allows researchers to make concrete estimates of the magnitude of residual terms of this model for any conceivable configuration of the source of light, massive bodies, and VLBI stations. The largest terms in the relativistic time delay which can affect the current VLBI observations are from the quadrupole and the angular momentum of the gravitating bodies that are known from the literature. These terms should be included in the new geodetic VLBI model for improving its consistency.

\vfill\eject

\ \vskip1truecm\noindent
{\Large\bf List of symbols}
\begin{itemize}

\item  [$G$:] universal gravitational constant
\item [$c$:] vacuum speed of light
\item [$M_\rA$:] mass of body A
\item [$M_L$:] $M_L = M_{i_1 \dots i_l}$, where each Cartesian index runs over $1,2,3$ or $x,y,z$. It denotes the
Cartesian mass-multipole moments of a  of degree $l$ (e.g., $M_{ij}$ denotes the Cartesian mass-quadrupole moments of a body)
\item [$(t,\vx)$:] time and spatial coordinates in the global reference system; $t =\,$TCB in the BCRS
\item [$x^0$:] $x^0 = ct$
\item [$x^i$:] $x^i = (x,y,z)$
\item [$(T,\vX)$:] time and spatial coordinates in a local system; $T = $TCG in the GCRS
\item [$X^0$:] $X^0 = cT$
\item [$X^a$:] $X^a = (X,Y,Z)$
\item [$g_{\mu\nu}$:] components of the BCRS metric tensor
\item [$G_{\alpha\beta}$:] components of  the metric tensor in a local coordinate system (mostly the GCRS)
\item [$\eta_{\mu\nu}$:] $\eta_{\mu\nu} = {\rm diag} (-1,+1,+1,+1)$, the Minkowskian metric
\item [$w(t,\vx)$:] global gravito-electric potential, generalizing the Newtonian potential $U(t,\vx)$
\item [$w^i(t,\vx)$:] global gravito-magnetic potential
\item [$(W,W^a)$:] gravito-electric and magnetic GCRS potentials
\item [$\delta_{ij}$:] $\delta_{ij} = 1$ if $i=j$; zero otherwise
\item [$\vz_\rA(t)$:] barycentric coordinate position of body A as function of global coordinate time $t$
\item [$\vz_\rA(T)$:] barycentric coordinate position of body A as function of local time $T$
\item [$(\wbar,\wbar^i)$:] external metric potentials
\item [TCB:] Barycentric Coordinate Time
\item [TCG:] Geocentric Coordinate Time
\item [TT:] Terrestrial Time
\item [TDB:] Barycentric Dynamical Time
\item [$L_G$:] a defining constant; defines TT in terms of TCG
\item [$L_B$:] a defining constant; defines TDB in terms of TCB
\item [$x^i_\rL(t)$:] light-ray trajectory in global coordinates; often the index $\rL$ is suppressed
\item [$\dot x^i$:] abbreviation for $dx_\rL^i(t)/dt$
\end{itemize}

\section{Introduction}
Very Long Baseline Interferometry (VLBI) is a very remarkable observational and measuring technique. Signals from radio sources such as quasars, located near
the edge of our visible universe are recorded by two or more radio antennas and the cross-correlation function between each pair of signals is constructed that leads
to the basic observable: the geometric time delay between the arrival times of a certain feature in the signal at two antennas. From this a wealth of information is deduced:
the positions and time and frequency dependent structure of the radio sources, a precise radio catalogue that presently defines the International Celestial Reference Frame
(ICRF) with an overall precision of about $40\,\mu$as for the position of individual sources and $10\,\mu$as for the axis orientation of the ICRF-2  (Jacobs et al.\ 2013).
In addition to information derived with other geodetic space techniques such as Satellite Laser Ranging (SLR) and Global Navigation Satellite Systems (GPS, GLONASS, GALILEO, BEIDOU)
it provides important information for the International Terrestrial Reference System (ITRF) with accuracies in the mm range.

VLBI is employed for a precise determination of Earth's orientation parameters related with precession-nutation, length of day and polar motion thus
providing detailed information about the various subsystems of the Earth (elastic Earth, fluid outer core, solid inner core, atmosphere, ocean, continental
hydrology, cryosphere etc.) and their physical interactions. In this way VLBI not only contributes significantly to  geophysics but also presents an important
tool to study  our environment on a global scale and its change with time.

To utilize the full power of VLBI, the establishment of a VLBI model with adequate precision is essential; at present such a model should have an internal
precision below $1\,$ps for the delay between the times of arrival of a radio signal at two VLBI stations separated by a continental baseline. Any reasonable VLBI model for Earthbound antennas has to describe a variety of different effects:
\begin{itemize}
\item [(1)] the propagation of radio signals from the radio sources to the antennas,
\item [(2)] the propagation of radio signals through the solar corona, planetary magnetospheres and interstellar medium,
\item [(3)] the propagation through the Earth's ionosphere,
\item [(4)] the propagation through the Earth's troposphere,
\item [(5)] the relation between the ICRS (International Celestial Reference System (better: GCRS (Geocentric Celestial Reference System)) and the ITRS (International Terrestrial Reference System),
\item [(6)] the time-dependent motion of antenna reference points in the ITRS,
\item [(7)] instrumental time-delays,
\item [(8)] clock instabilities.
\end{itemize}

In this article we will focus on the first issue. Effects from the signal propagation through the troposphere are included in the model.  Present VLBI tries to reach mm accuracies so the underlying theoretical model should have an accuracy of better than $0.3\,$ps. This number has to be compared with the largest relativistic terms; e.g., the gravitational time delay near the limb of the Sun amounts to about $170\,$ns for a baseline of $6000\,$km. So at the required level of accuracy the model has to be formulated within the framework of Einstein's theory of gravity.

The standard reference to such a relativistic VLBI-model are the IERS Conventions
2010 (IERS Technical Note No.\ 36, G.Petit, B.Luzum (eds.)). As explained
there the IERS-model is based upon a consensus model (not necessarily intrinsically consistent) as  described in Eubanks
(1991). The consensus model was based upon a variety of relativistic VLBI models with accuracies in the picosecond range.

The purpose of the present  paper is first to re-derive the consensus  model for Earthbound baselines within a more consistent framework. Then we  extend and improve this formalism.
With a few exceptions, e.g.\ for the tropospheric delay, all results are derived explicitly using a well accepted formulation of relativistic celestial mechanics.
The paper tries to be as detailed as possible. This will be of help for the reading of non-experts but also for further theoretical
work on the subject. The paper basically confirms the expressions from the consensus model. In several respects, however, we go beyond the standard model. E.g.\ terms related with the acceleration of the Earth might become interesting at the level of a few femtoseconds (fs) for baselines of order $6000\,$km; they grow quadratically with the station distance to the geocenter.
In the gravitational time delay we consider the gravitational field of a moving body with arbitrary mass- and spin-multipole moments.
Another point is the parallax expansion for radio sources at finite distance which is treated with a new parallax expansion (Section 2).

We believe that our new formulation has an intrinsic accuracy of order $10\,$fs (femtoseconds), but further checks have to be made to confirm that statement.

The time delay in VLBI measurements is first formulated in the Barycentric Celestial Reference System (BCRS) where the signal propagation from the radio source to the antennas is described; at this place BCRS baselines $\vb$ are introduced. Then
the basic time delay equation (\ref{geom-transformed}) is derived from the Damour-Soffel-Xu (1991) formulation of relativistic reference systems. It provides the transformation formulas from the BCRS to the GCRS where GCRS baselines $\vB$ are defined. In this  basic time delay equation only the gravitational (Shapiro) time delay term is not written out explicitly. In Appendix
C the Shapiro term is treated exhaustively.

The organization of this article is as follows: Section 2 contains the main part of the paper where all central results can be found. In Section 3 some conclusions are presented.
All technical details and derivations of results can be found in the Appendices.

Appendix A presents relevant parts of the theory of relativistic reference systems where the transformations between the BCRS and the GCRS are discussed in detail.

Appendix B discusses the form of the metric tensor for the solar system at the first and second post-Newtonian level.

Appendix C focuses on the gravitational time delay in the propagation of electromagnetic signals or light-rays. In the BCRS the post-Newtonian equation of a light-ray (at various places we drop the index L referring to light-ray) takes the form
\begin{equation}\label{eqone}
\vx_\rL(t) = \vx_0 - c \vk (t - t_0) + \vx^{\rm G}(t) \equiv \vx^{\rm N}(t) + \vx^{\rm G}(t) \, ,
\end{equation}
where $\vn = - \vk$ is a Euclidean unit vector ($n^i n^i = 1$) in the direction of light-ray propagation.
I.e.\ to the Newtonian form of the light-ray trajectory,
\begin{equation}
\vx_\rL^\rN(t) = \vx_0 - c \vk (t - t_0)
\end{equation}
one adds
a post-Newtonian  term proportional to $1/c^2$ that is determined by the gravitational action of the solar system bodies (the gravitational light-deflection and  the gravitational time-delay (Shapiro)). In this Appendix results for the Shapiro term can be found for a (moving) gravitating body with arbitrary mass- and spin-multipole moments. Here technically the so-called Time Transfer Function (TTF) is employed.

Finally Appendix D provides additional derivations of certain  statements of the main section.


%
%
\section{An advanced relativistic VLBI model for geodesy}

Since this article concentrates on Earthbound baselines it is obvious that at least two space-time reference systems
have to be employed:

\begin{itemize}
\item [(i)] One global coordinate system $(t,x^i)$, in which the light propagation from remote sources (e.g.\ a quasar)
can be formulated and the motion of solar system bodies can be described. The
origin of this system of coordinates will be chosen as the barycenter of the solar system,
thus our global system will be {the} Barycentric Celestial Reference System
(BCRS). Its time coordinate will be TCB (Barycentric Coordinate Time).

\item[(ii)] Some geocentric coordinate system $(T,X^a)$, comoving with the Earth, in which geodetically meaningful baselines can be defined.
We will employ the Geocentric Celestial Reference System (GCRS) to this end with $T = \,$TCG (Geocentric Coordinate Time) as basic timescale.
\end{itemize}

One might employ additional reference systems for a highly-accurate VLBI model. One might introduce  topocentric reference systems but they will not be needed in the following. On might introduce some galacto-centric celestial reference system; but since the problem of galactic rotation will not be touched (e.g.\ Lambert 2011, Titov et al 2011) this also will not be needed. One might modify the BCRS to account for the Hubble expansion of the universe; an attempt in this direction can be found e.g.\ in Klioner \& Soffel 2004. There it was shown that if the generalized BCRS coordinates are chosen properly "effects" from the Hubble expansion on planetary orbits and the propagation of light-rays are completely negligible in the solar system.

Barycentric Coodinate Time, TCB, and Geocentric Coordinate Time, TCG, are the fundamental time coordinates of the BCRS and the GCRS, respectively.
The relationship between them, according to (\ref{time-trans}) is given by
\begin{equation}
{\rm TCB} - {\rm TCG} = c^{-2} \left[ \int_{t_0}^t \left( {\vv_\rE^2 \over 2} + \wbar(\vz_\rE) \right) dt
+ v_\rE^i r_\rE^i \right] + \cO(c^{-4}) \, ,
\end{equation}
with $r_\rE^i = x^i - z_\rE^i$.

Note that no real clock on Earth will show directly TCG. Real atomic clocks on Earth define International Atomic Time,
TAI, that differs from Terrestrial Time, TT, only by a shift of $32.184\,$s.  According to an IAU-2000 resolution B1.9 Terrestrial time TT is defined by
\begin{equation} \label{TT}
{\rm TT} = {\rm TCG} - L_G \times ({\rm JD}_{\rm TCG}  - 2443144.5003725) \times 86400 \, ,
\end{equation}
where ${\rm JD}_{\rm TCG}$ is TCG-time expressed as Julian date. $L_G$ is a defining constant with
\begin{equation} \label{LG}
L_G = 1 - {d({\rm TT}) \over d({\rm TCG})} =  6.969290134 \times 10^{-10} \, .
\end{equation}

For the use in ephemerides the time scale TDB (Barycentric Dynamical Time) was introduced. IAU resolution 3 of 2006 defines TDB as a linear transformation
of TCB. As of the beginning of 2011, the difference between TDB and TCB was about 16.6 seconds. TDB is defined by (e.g.\ Soffel at al.\ 2003)
\begin{equation}
{\rm TDB} = {\rm TCB} - L_B \times ({\rm JD}_{\rm TCB} - T_0) \times 86400 + {\rm TDB}_0 \, ,
\end{equation}
\begin{eqnarray*}
L_B &=& 1.550519768 \times 10^{-8} \, , \\
{\rm TDB}_0 &=& - 6.55 \times 10^{-5} \, {\rm s} \, , \\
T_0 &=& 2443144.5003725 \, .
\end{eqnarray*}

\begin{figure}
\begin{center}
\vspace{-50mm}
\includegraphics[scale=0.5]{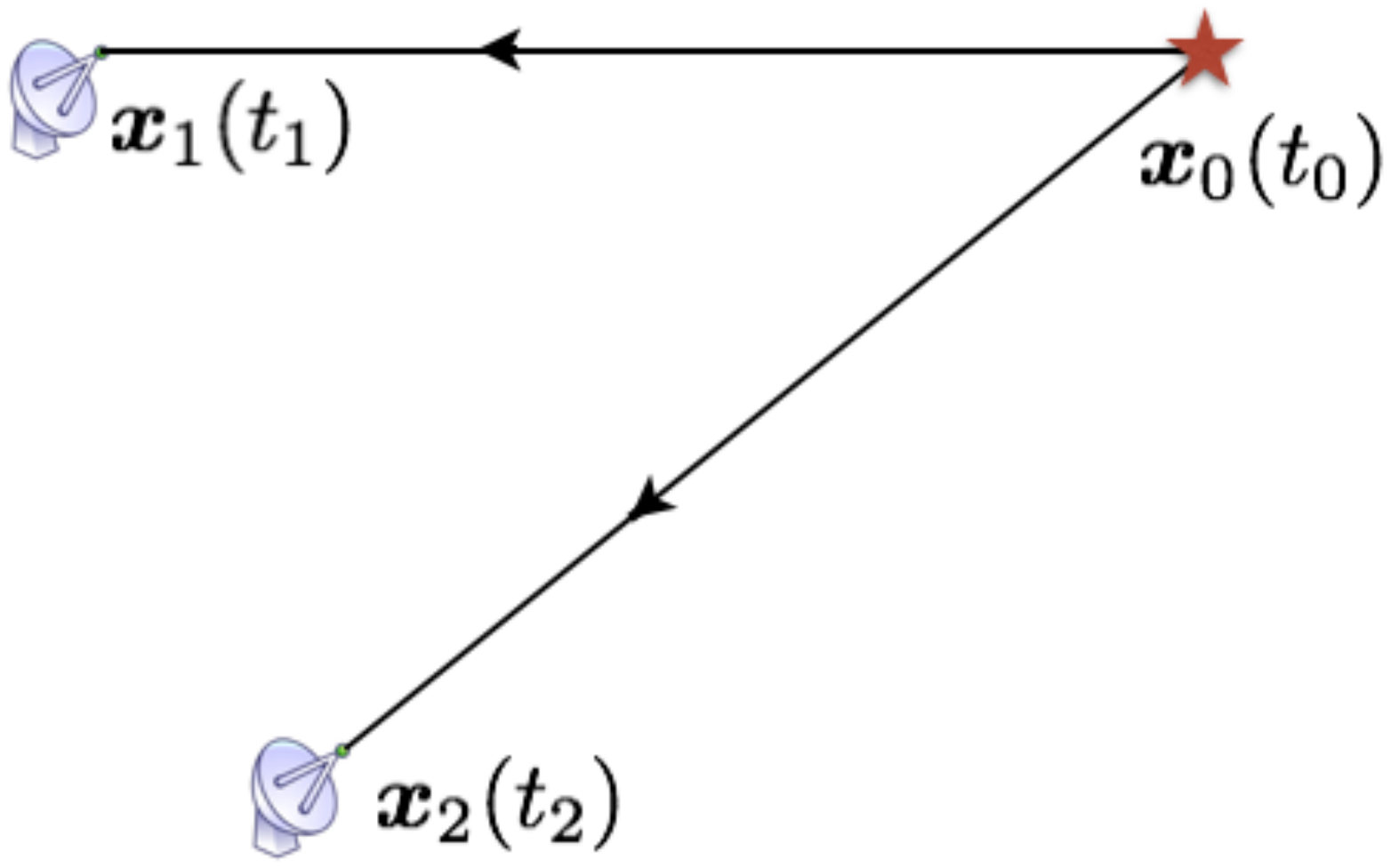}%
\vspace{-40mm}
\caption{\label{fig:vlbischeme} Geometry in the problem of an elementary VLBI
measurement.}
\end{center}
\end{figure}

Due to the Earth's acceleration the GCRS is only a {\it local} reference system, i.e., its spatial coordinates do not extend to infinity (e.g.\ Misner et al.\ 1973).
For that reason the signal propagation from a sufficiently remote radio source to the antennae has to be formulated in the BCRS.
For the problem of propagation  times we consider two light-rays, both
originating from a source at BCRS position $\vx_0$ and time $t_0$ (see
Fig.\ref{fig:vlbischeme}).
Each of these two light-rays is described in BCRS coordinates by an equation
of the form
\begin{equation}\label{pupu}
\vx_\rL^{(i)}(t) = \vx_0 - c \vk_i (t - t_0) + \vx^{\rm G}_i(t) \equiv \vx^{\rm N}_i(t) + \vx^{\rm G}_i(t)  \, ,
\end{equation}
where the Euclidean unit vector $\vk_i$ points from antenna $i$ towards the
radio source,
\begin{equation*}
\vk_i = {\vx_0 - \vx_i \over \vert \vx_0 - \vx_i \vert } \, .
\end{equation*}
We now assume that light-ray number $i$ ($i = 1,2$) reaches antenna $i$ at
barycentric coordinate position $\vx_i$ at barycentric coordinate time $t_i$,
so that
\begin{equation}
\vx_\rL^{(i)}(t_i) = \vx_i(t_i) \, .
\end{equation}
From (\ref{pupu}) and including influences of the atmosphere, we then get
\begin{equation}
\Delta t \equiv t_2 - t_1  =  (\Delta t)_{\rm geom} + (\Delta t)_{\rm grav} + { (\Delta t)_{\rm atm}}
\end{equation}
with
\begin{eqnarray}
(\Delta t)_{\rm geom} &=& -{1 \over c} \vk_2 \cdot (\vx_2(t_2) - \vx_0)  + {1
\over c} \vk_1 \cdot ({\vx_1(t_1)} - \vx_0) \\
(\Delta t)_{\rm grav} &=&  +{1 \over c} \vk_2 \cdot (\vx^{\rm G}_2(t_2)) -
{1 \over c} \vk_1 \cdot (\vx^{\rm G}_1(t_1)) \\
(\Delta t)_{\rm atm} &=& \delta t_{{\rm atm}_2} - \delta t_{{\rm atm}_1}  \, .
\end{eqnarray}
From Kopeikin \& Han (2015), the atmospheric delay can be written as
\begin{equation}
\delta t_{{\rm atm}_i} = \int_{t_{ai}}^{t_i} {(n-1)\left(1+\frac{2}{c}\vk\cdot\vv_{\rm atm}\right)dt} \, , \label{delt-atm}
\end{equation}
where $t_{ai}$ is  TCB time when the light ray enters the atmosphere, $n$ the index of refraction of the troposphere and $\vv_{\rm atm}$ the BCRS velocity of some  tropospheric element on the path of the signal's propagation.
\subsection{Very remote radio sources}

\subsubsection{Barycentric model}

We will consider very remote sources first, so that we can neglect the parallaxes; for them
 we can put $\vk_1 = \vk_2 = \vk$ so that
\begin{equation} \label{veck}
\vk = {\vx_0 \over \vert \vx_0 \vert} \, ,
\end{equation}
and
\begin{eqnarray}\label{delt-remote}
 (\Delta t)_{\rm geom} &=& -{1
\over c} \vk \cdot (\vx_2(t_2)   -  \vx_1(t_1)) \label{delt-geom} \\
(\Delta t)_{\rm grav} &=& +{1
\over c} \vk \cdot (\vx^{\rm G}_2(t_2)   -  \vx^{\rm G}_1(t_1)) \, . \label{delt-grav}
\end{eqnarray}

Let us define baselines at signal arrival time $t_1$ at antenna 1. Let the
barycentric baseline $\vb$ be defined as
\begin{equation}\label{BCRSbaseline}
\vb \equiv \vb(t_1) \equiv \vx_2(t_1) - \vx_1(t_1) \, ,
\end{equation}
then a Taylor expansion of $\vx_2(t_2)$ in (\ref{delt-geom})  about $t_1$ yields
\begin{equation}\label{VLBIbary}
(\Delta t)_{\rm geom} =\  -{1 \over c} (\vb \cdot \vk) \left( 1 - {1 \over c}(\dot\vx_2
\cdot \vk) + {1 \over c^2} (\dot \vx_2 \cdot \vk)^2 \right.   - \left. {1
\over 2 c^2} (\vb \cdot \vk) (\ddot  \vx_2 \cdot \vk) \right)  + \cO(c^{-4}) \, ,
\end{equation}
all quantities now referring to TCB $t_1$.

\subsubsection{Geocentric baselines}
Clearly for Earth-bound baselines we want to define them in the
GCRS. Let us define a {GCRS} baseline via
\begin{equation}\label{GCRSbaseline}
\vB \equiv {\vX_2(T_1) - \vX_1(T_1)} \, .
\end{equation}
By using the coordinate transformations between barycentric and geocentric spatial coordinates (resulting from Lorentz-contractions terms and  corresponding terms related with gravitational potentials and acceleration terms of the geocenter) and time coordinates (resulting from time dilation and gravitational redshift terms) one finds a relation between a barycentric baseline $\vb$ and the corresponding geocentric one $\vB$ (the baseline equation (\ref{baselineeq})).

Let $\Delta t = t_2 - t_1$ and $\Delta T = T_2 - T_1$ be the coordinate time difference between signal arrival times at antenna $2$ and $1$ in the
BCRS and in the GCRS, respectively.
A detailed analysis of the time transformation then leads to a relation between $\Delta t$ and $\Delta T$ (relation (\ref{deltateq}) from Appendix D).
Using this relation  we get a delay equation of the form
\begin{eqnarray} \label{geom-transformed}
\Delta T &=& - {1 \over c} (\vB \cdot \vk) \nonumber \\
&& + {1 \over c^2} (\vB \cdot \vk) (\vk \cdot \vv_2) - {1 \over c^2} (\vB \cdot \vv_\rE)  \nonumber \\
&& + {1 \over c^3} (\vB \cdot \vk) \left[ (\vv_\rE \cdot \vV_2) - (\vk \cdot \vv_2)^2 + 2 \wbar (\vz_\rE) + {1 \over 2} \vv_\rE^2 \right. \nonumber \\
&&
\ \ \ \ \ \ \ \ \ \ \ \ \ \ \ \ \ \ - \left. {1 \over 2} (\vB \cdot \vk) (\vk \cdot \va_2) + (\va_\rE \cdot \vX_2) \right]  \nonumber \\
&& + {1 \over c^3} (\vB \cdot \vv_\rE) \left[ (\vk \cdot \vV_2) + {1 \over 2} (\vk \cdot \vv_\rE) \right]  \nonumber \\
&& - {1 \over c} \vk \cdot \Delta \vxi + {\frac{\Delta t_{\rm grav}}{1+\vk \cdot \vv_2/c}} \nonumber \\
&& + { (\delta T_{{\rm atm}_2} - \delta T_{{\rm atm}_1}) + \delta T_{{\rm atm}_1}\frac{\vk\cdot(\vV_2-\vV_1)}{c}} \, .
\end{eqnarray}
with ($\vxi$ is defined in (\ref{xieq}) of Appendix A)
\begin{eqnarray}
\Delta\vxi &=& \vxi(T_1,\vX_2) - \vxi(T_1, \vX_1)  \nonumber \\
&=&
{1 \over c^2} \left[ {1 \over 2} \va_\rE (\vB \cdot (\vX_1 + \vX_2)) - \vX_2 (\va_\rE \cdot \vX_2)
+ \vX_1 (\va_\rE \cdot \vX_1) \right] \, .
\end{eqnarray}
In this basic time delay equation $\vB$ is the geocentric baseline from (\ref{GCRSbaseline}), $\vk$ the Euclidean unit vector from the barycenter to the radio source from (\ref{veck}),
$\vv_2$ the barycentric coordinate velocity of antenna 2, $\vV_2$  the corresponding geocentric velocity (to Newtonian order $\vv_2 = \vv_\rE + \vV_2$), $\wbar(\vz_\rE)$ the external gravitational potential resulting from all solar system bodies except the Earth taken at the geocenter, $\vv_\rE$ and $\va_\rE$ are the BCRS velocity and acceleration of the geocenter and $\vX_i$ is the GCRS coordinate position of antenna $i$.
The atmospheric terms can be derived to sufficient accuracy from
\begin{equation}
\delta T_{\rm atm} = {\delta t_{\rm atm} \over 1 + \frac 1c \vk \cdot \vv_2} \, .
\end{equation}
Explicit expressions for  $\Delta t_{\rm grav}$ are given below.
In Appendic D it is shown that the basic time delay equation (\ref{geom-transformed}) can be derived directly without the introduction of some BCRS baseline.

A comparison of (\ref{geom-transformed}) with expression (11.9) from the IERS Conventions
shows that all terms from the Conventions are contained in the basic time delay equation after an expansion in terms of $1/c$. The $\vk \cdot \Delta \vxi$-term is missing in the Conventions since for earthbound baselines the order of magnitude is of order a few fs; note that this term grows quadratically with the station distance to the geocenter (this term is known from the literature; Soffel et al.\ 1991).

\subsubsection{Scaling problems}
Our baseline $\vB$ was defined by a difference of spatial coordinates in the
GCRS, i.e., it is related with TCG, the basic GCRS timescale. In modern
language of the IAU Resolutions (e.g., IERS Technical Note No.\ 36) our $B$ is TCG-compatible, $\vB = \vB_{\rm TCG}$.

We will assume that the station clocks are synchronized to UTC, i.e., there
rates are TT-compatibel. The geocentric space coordinates resulting from a
direct VLBI analysis, $\vX_{\rm VLBI} = \vX_{\rm TT}$, are therefore also
TT-compatible. According to (\ref{LG})
the TRS space coordinates recommended by IAU and IUGG resolutions, $\vX_{\rm
TCG}$, may be obtained {\it a posteriori} by
\begin{equation}
\vX = \vX_{\rm TCG} = {\vX_{\rm VLBI} \over  1 - L_G} \, .
\end{equation}

\subsection{The gravitational time delay in VLBI}

The gravitational time delay or Shapiro effect for a single light-ray is discussed extensively in Appendix C. In this Appendix it is treated with the method of the Time Transfer Function (TTF)
defined by
\begin{equation}
\cT(t_0,\vx_0;\vx) = t - t_0 \, .
\end{equation}
Here it is assumed that a light-ray starts from coordinate position $\vx_0$ at coordinate time $t_0$ and reaches the point $\vx$ at time $t$. We had assumed that such a light-ray reaches antenna $i$ at BCRS position $\vx_i$ at TCB $t_i$ so that
\begin{equation}
t_i - t_0 = \cT(t_0,\vx_0; \vx_i) \, .
\end{equation}

The VLBI gravitational time delay is just a differential delay, as time difference in the arrival time of a signal at the two radio
antennas:
\begin{equation}\label{gravtimedelay}
(\Delta t)_{\rm grav} = (t_2 - t_0) - (t_1 - t_0) = \cT(t_0,\vx_0;\vx_2) - \cT(t_0,\vx_0;\vx_1) \, .
\end{equation}
From this relation $(\Delta t)_{\rm grav}$ can be derived from the expressions given in Appendix C. The dominant terms resulting from the mass-monopole, mass-quadrupole and spin-dipole of solar system bodies are given explicitly in the next subsections; they are already known from the literature.

\subsubsection{Mass-monopoles to 1PN order at rest}

Let us first consider the Sun at $\vx_{\rm S} = 0$ (moving bodies are considered in the next Subsection).
From (\ref{mass-monopole}) we get:
\begin{equation}
(\Delta t)^{\rm Sun}_{\rm pN}  = {2 G M_{\rm S} \over c^3} \left[ \ln
\left( {\vert \vx_2\vert - \vx_2 \cdot \vk_2  \over \vert \vx_1 \vert - \vx_1
\cdot \vk_1 } \right) + \ln \left( {\vert \vx_0\vert - \vx_0 \cdot \vk_1
\over \vert \vx_0 \vert - \vx_0 \cdot \vk_2 } \right) \right] \, .
\end{equation}
An expansion yields
\begin{equation}
\begin{split}
\vk_i =&\ \vk + {1 \over \vert \vx_0\vert} [\vk \cdot (\vx_i \cdot \vk) -
\vx_i] \\
&- {1 \over \vert \vx_0 \vert^2 } \left( \vx_i (\vx_i \cdot \vk) + {1 \over
2} \vk \vx_i^2 - {3 \over 2} \vk (\vx_i \cdot \vk)^2 \right) + \dots .
\end{split}
\end{equation}
Using this result we find:
\begin{equation*}
{\vert \vx_0\vert - \vx_0 \cdot \vk_1  \over \vert \vx_0 \vert -  \vx_0 \cdot
\vk_2} = {1 - \vk \cdot \vk_1 \over 1 - \vk \cdot \vk_2} = {\vx_1^2 - (\vx_1
\cdot \vk)^2 \over \vx_2^2 - (\vx_2 \cdot \vk)^2}
\end{equation*}
so that (Finkelstein et al.\ 1983; Soffel 1989)
\begin{equation}
(\Delta t)_{\rm M,pN}^{\rm Sun} = {2 G M_{\rm S} \over c^3} \ln \left(
{\vert\vx_1\vert + \vx_1 \cdot \vk \over  \vert\vx_2\vert + \vx_2 \cdot \vk}
\right) \, .
\end{equation}
The time difference $\Delta t$ can be neglected in the ln-term and writing
\begin{equation*}
\vx_i = \vx_{\rm E} + \vX_i \, ,
\end{equation*}
we obtain (Finkelstein et al.\ 1983; Zeller et al.\ 1986):

\begin{small}
\begin{equation}
(\Delta t)_{\rm M,pN}^{\rm Sun} = {2 G M_{\rm S} \over c^3} \ln \left( { r_{\rm
E}(1 + \ve_{\rm E} \cdot \vk) + \vX_1 \cdot (\ve_{\rm E} + \vk) + \vX_1^2/2
r_{\rm E} - (\ve_{\rm E} \cdot \vX_1)^2/2 r_{\rm E} \over r_{\rm E}(1 +
\ve_{\rm E} \cdot \vk) + \vX_2 \cdot (\ve_{\rm E} + \vk) + \vx_2^2/2 r_{\rm
E} - (\ve_{\rm E} \cdot  \vx_2)^2/2 r_{\rm E}} \right) \, ,
\end{equation}
\end{small}
with
\begin{equation*}
\ve_{\rm E} \equiv \vx_{\rm E}/r_{\rm E} \, , \quad r_{\rm E} = \vert
\vx_{\rm E} \vert = (x_{\rm E}^i x_{\rm E}^i)^{1/2} \, .
\end{equation*}

\medskip
Next we consider some planet A at rest in the BCRS. The corresponding time
delay is then given by
\begin{equation}
(\Delta t)^{\rm planet\ A}_{\rm M,pN} = {2 G M_{\rm A} \over c^3} \ln \left(
{\vert\vr_{\rm A1}\vert + \vr_{\rm A1} \cdot \vk \over  \vert\vr_{\rm
A2}\vert + \vr_{\rm A2} \cdot \vk} \right) \, ,
\end{equation}
where
\begin{equation*}
\vr_{\rm Ai} \equiv \vx_i(t_i) - \vx_{\rm A} \, .
\end{equation*}
For the gravitational time delay due to the Earth one finds
\begin{equation}
(\Delta t)_{\rm M,pN}^{\rm Earth} = {2 G M_{\rm E} \over c^3} \ln \left(
{\vert\vX_1\vert + \vX_1 \cdot \vk \over \vert\vX_2\vert +  \vX_2 \cdot \vk}
\right) \, ,
\end{equation}
if the motion of the Earth during signal propagation is neglected.

Note that the maximal gravitational time delays due to Jupiter, Saturn,
Uranus and Neptune are of order 1.6(Jup), 0.6(Sat), 0.2(U), and 0.2(N)
nanosec, respectively, but these values decrease rapidly with increasing
angular distance from the limb of the planet (Klioner 1991). E.g., 10 arcmin from the center
of the planet the gravitational time delay amounts only to about 60 ps for
Jupiter, 9 ps for Saturn, and about 1 ps for Uranus.

\subsubsection{Mass-monopoles to 1PN order in motion}

If the motion of a gravitational body A, say a planet in the solar system,  is considered, we face several
problems (Kopeikin 1990; Klioner 1991; Klioner 2003). One is the instant of time when the position of the
massive body A should be taken in the equation of the time delay. {According to Kopeikin (1990) and Klioner (1991) the errors are
minimized if the moment of closest approach of the unperturbed light ray to the body A is taken.}  {Kopeikin \& Sch\"afer (1999) proved the time at which the body is taken on its orbit in the time delay equation is the retarded time while the time of the closest approach is an approximation. The difference between the two instants of time is practically small but important from the principal point of view, in the physical interpretation of  time-delay experiments. }
We had written the unperturbed light-ray in the form
$ \vx_\rL^{\rm N}(t) = \vx_0 - c \vk (t - t_0)\, .$
Because the light rays moving from the source of light to each VLBI station are different
we define the impact parameter vector of each light ray with respect to  body A as follows (Kopeikin \& Sch\"afer 1999):
\begin{equation}
\label{tde2356}
\vd_{{\rm A}i} =  \vk \times {(\vr_{{\rm A}i} \times \vk)}
\end{equation}
with
\begin{equation}
\label{eq9634}
\vr_{{\rm A}i} \equiv \vx_i(t_i) - \vx_{\rm A}(t_{{\rm A}i}) \, ,
\end{equation}
{where $t_{{\rm A}i}$ is the retarded time}
\begin{equation}
\label{est452}
t_{{\rm A}i} = t_i - {r_{{\rm A}i} \over c} \, .
\end{equation}
The gravitational time delay in the time of arrivals of two light rays at two VLBI stations resulting from body
A was given by Kopeikin \& Sch\"afer (1999) and has the following form 
\begin{equation}\label{monopole-move}
(\Delta t)^{\rm A}_{{\rm M},l=0} = {2 G M_{\rm A} \over c^3} \left[ 1 + \vk \cdot \vbeta_{\rm A}{(t_{{\rm A}1})} \right]
 \ln \left(
{\vert\vr_{\rm A1}\vert + \vr_{\rm A1} \cdot \vk \over \vert\vr_{\rm A2}\vert
+ \vr_{\rm A2} \cdot \vk} \right) \, ,
\end{equation}
where $\vr_{{\rm A}1}$ and $\vr_{{\rm A}2}$ are to be taken from \eqref{eq9634} with the retarded times $t_{{\rm A}1}$ and $t_{{\rm A}2}$ calculated from (\ref{est452}) for $i=1,2$ respectively.
The time delay \eqref{monopole-move}  has the same form as (\ref{moving-mass}) of Appendix C for the case of the body A moving with a constant velocity (Kopeikin 1997; Klioner \& Kopeikin 1992).
Klioner (1991) has estimated the effects from the translational motion of gravitating bodies. For an earthbound baseline of $6000\,$km the additional effect near the limb of the Sun amounts to $0.01\,$ps, of Jupiter $0.07\,$ps and of Saturn $0.02\,$ps.

\subsubsection{The influence of mass-quadrupole moments} The gravitational time delay due to the mass-quadrupole moment of body
A can be described by
\begin{equation}
(\Delta t)_{{\rm M},l=2} = {G \over c^3} M^{\rm A}_{pq} (f_{{\rm A}2}^{pq} -
f_{{\rm A}1}^{pq})
\end{equation}
with
\begin{equation}
f_{{\rm A}i}^{pq} = (1 - (\vk \cdot \vn_{{\rm A}i})^3)\, {k^p k^q \over
d_{{\rm A}i}^2} + {2 k^p d_{{\rm A}i}^q \over r_{{\rm A}i}^3} + (2 - 3 \vk
\cdot \vn_{{\rm A}i} + (\vk \cdot \vn_{{\rm A}i})^3)\, {d_{{\rm A}i}^p
d_{{\rm A}i}^q \over d_{{\rm A}i}^4} \, .
\end{equation}
Here,
\begin{equation*}
\vn_{{\rm A}i} \equiv \vr_{{\rm A}i}/r_{{\rm A}i}
\end{equation*}
and $\vd_{{\rm A}i} = \vn_i \times (\vr_{{\rm A}i} \times \vn_i)$ (Kopeikin 1997; Klioner \& Kopeikin 1992).
Maximal effects from the oblateness of gravitating bodies for $b \sim 6000\,$km are of order $0.2\,$ps for the Sun, $21\,$ps for Jupiter, $8\,$ps for Saturn, $2\,$ps for Uranus and $0.7\,$ps for Neptune (Klioner 1991).

\subsubsection{The influence of higher mass-multipole moments}

In Appendix C we present all necessary formulas to compute the gravitational time delay due to higher mass-multipole moments (potential coefficients with $l > 2$). For the Sun there are indications that the $J_4$ term is surprisingly large, only a factor of ten smaller than $J_2$ (Ulrich \& Hawkins 1980). This implies that very close to the limb of the Sun the $J_4$-term might lead to a time delay as large as $0.02\,$ps. More detailed studies are needed to better estimate the $J_4$-effect from the Sun. For Jupiter $J_4$ is about $-587 \times 10^{-6}$, roughly a factor of 25 smaller than $J_2 = 14696 \times 10^{-6}$ (Jacobsen 2003) so the maximal time delay might be slighly less than $1\,$ps. Note that the gravitational field of a body produced by its hexadecupole falls of much faster with distance from the body than the quadrupole field. So for real geodetic VLBI observations such hexadecupole effects might be smaller than a fs and hence negligibe.

\subsubsection{The influence of spin-dipole moments} The gravitational time
delay due to the spin-dipole moment of body A can be obtained {from (\ref{spin-dipole})} as difference for the two antennas.
Using $\vd_{\rA i} = \vr_{\rA i } - \vk (\vr_{\rA i} \cdot \vk)$ and $d^2_{\rA i} = (r_{\rA i} + \vk \cdot \vr_{\rA i})
(r_{\rA i} - \vk \cdot \vr_{\rA i})$ one finds ((4.11) of Klioner, 1991; Kopeikin \& Mashhoon 2002)
\begin{equation}
(\Delta t)_{S_a} = {2 G \over c^4} (\vk \times \vS_{\rm A}) \cdot( \vF_{{\rm
A}2} - \vF_{{\rm A}1})\, , \qquad \vF_{{\rm A}i} \equiv {\vn_{{\rm A}i} \over
r_{{\rm A}i} + \vk \cdot \vr_{{\rm A}i} } \, .
\end{equation}
Spin-dipole effects for $b \sim 6000\,$km near the limb of the rotating body are of order $0.06\,$ps for the Sun, and $0.02\,$ps for Jupiter (Klioner 1991). Effects from higher spin-moments with $l > 4$ are even smaller (see e.g.\ Meichsner \& Soffel 2015 for related material).

\subsubsection{2PN mass-monopoles at rest}
From Klioner (1991) (see also Brumberg 1987)
 we get the gravitational time delay from a mass-monopole
A to post-post Newtonian order in the form
\begin{equation}
\begin{split}
(\Delta t)_{\rm M,ppN} = &{G^2 M_{\rm A}^2 \over c^5} \left[ - {4 \over
r_{\rA 2} + \vk \cdot \vr_{\rA 2}} + {4 \over r_{\rA 1} + \vk \cdot \vr_{\rA
1}} + {\vk \cdot \vn_{\rA 2} \over 4 r_{\rA 2}} - {\vk \cdot \vn_{\rA 1}
\over 4
r_{\rA 1}}  \right. \\
& + \left. {15 \over 4\vert \vk \times \vr_{\rA 2}\vert} \arccos (\vk\cdot
\vn_{\rA 2}) - {15 \over 4\vert \vk \times \vr_{\rA 1}\vert} \arccos (\vk
\cdot \vn_{\rA 1}) \right] \, .
\end{split}
\end{equation}
The first two terms are the dominant ones and a further expansion of these
two terms leads to expression (11.14) in the IERS-2010 (Richter and Matzner,
1983; Hellings, 1986). Maximal time delays from 2PN effects ($b \sim 6000\,$km) are of order $307\,$ps for the Sun, $1.5\,$ps for Jupiter, $0.4\,$ps for Saturn, $0.1\,$ps for Uranus and
$0.3\,$ps for Neptune (Klioner 1991).

\subsection{Radio sources at finite distance}

Let us now consider the case of a radio source at finite distance. The vacuum part of the time-delay is
\begin{equation}\label{vacvac}
(\Delta t)_{\rm v} = t_{{\rm v}_2} - t_{{\rm v}_1} = {\vert \vx_2(t_2) - \vx_0 \vert \over c} - {\vert \vx_1(t_1) - \vx_0 \vert \over c} + \Delta t_{\rm grav} \, ,
\end{equation}
where ${\vx}_0$ is the coordinate of the radio source taken at the time of
emission: ${\vx}_0={\vx}_0(t_0)$, and ${\vx}_1$, ${\vx}_2$ are the spatial
coordinates of the first and second VLBI stations taken at the times $t_1$
and $t_2$ respectively. A geometric demonstration of these coordinates and
corresponding vectors are shown in Fig.\ 2 and Fig.\ 3.
\begin{figure}[h!]
\begin{center}
\vspace{-50mm}
\includegraphics[scale=0.5]{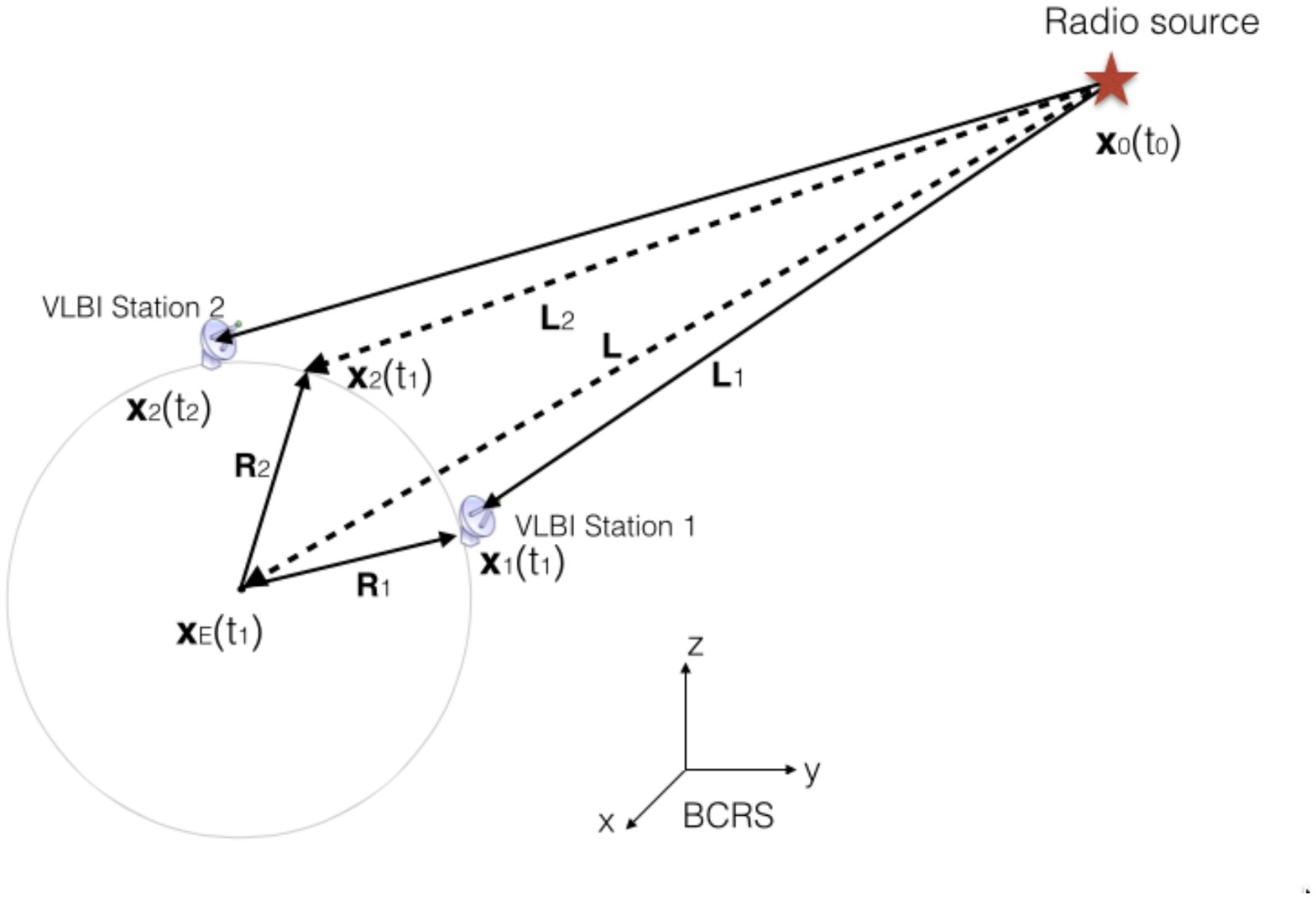}
\vspace{-40mm}
\caption{Geometry in the problem of a VLBI observation of an
object at finite distance.}\label{vlbifig1}
\end{center}
\end{figure}
\begin{figure}[h!]
\begin{center}
\vspace{-20mm}
\includegraphics[scale=0.5]{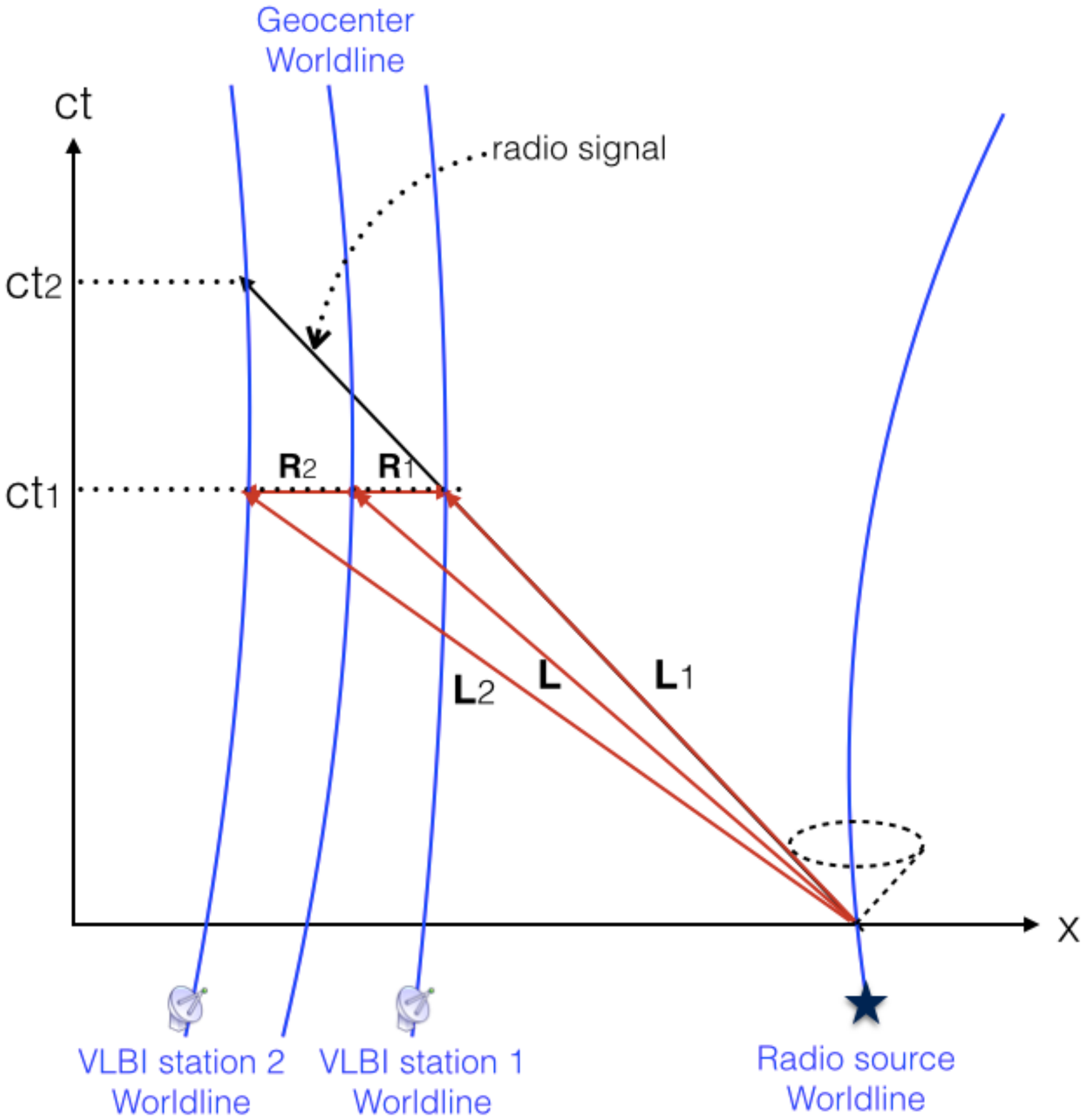}
\vspace{-20mm}
\caption{A spacetime diagram of the VLBI observation of a close object.}\label{closevlbifig2}
\end{center}
\end{figure}

Coordinates of all VLBI stations should be referred to the time of reception of the radio signal at the clock of the first VLBI station which is considered as the primary time reference.

Let us introduce the vectors
\begin{equation}
\vL_2 \equiv  \vx_2(t_1) - \vx_0 \, , \qquad  \vL_1 \equiv  \vx_1(t_1) - \vx_0 \, ,
\end{equation}
then in Appendix D it is shown that the vacuum part of the time-delay to sufficient accuracy can be written in the form
\begin{eqnarray} \label{finitedisdelt}
(\Delta t)_{\rm v}  &=& (\Delta t_0 + \Delta t_{\rm grav}) \left[ 1 - c^{-1} \vk_2 \cdot \dot\vx_2 + c^{-2} (\vk_2 \cdot \dot \vx_2)^2 - c^{-3}(\vk_2 \cdot \dot \vx_2)^3\right] \nonumber \\
&& - \frac 12 c^{-1} \vk_2 \cdot \ddot \vx_2 (\Delta t_0)^2 + \frac 12 c^{-1} L_2^{-1} \vert \vk_2 \times \dot \vx_2 \vert^2 (\Delta t_0)^2  \label{closegeomdelay2}
\end{eqnarray}
with
\begin{equation}
\Delta t_0 \equiv {L_2 - L_1 \over c}
\end{equation}
and
\begin{equation}
\vk_2 \equiv - {\vL_2 \over L_2} \, .
\end{equation}
We omit $\Delta t_\text{grav}$ in the quadric term because of $\Delta t_\text{grav}  \ll \Delta t_0$. Equation \eqref{closegeomdelay2} is sufficient for processing VLBI observation with the precision about 10 fs level. Sekido and Fukushima (Sekido \& Fukushima 2006)
used the Halley's method to solve the quadratic equation \eqref{geomtime3}. Their result is fully consistent with our (approximate) solution \eqref{closegeomdelay2}.
In (\ref{finitedisdelt}) the two vectors
$\vL_1$ and $\vL_2$, are employed.
These vectors are directed from the radio source to the first and second VLBI
stations respectively and cannot be calculated directly in practical work.
Instead, a decomposition in two vectors is used. More specifically,
\be\la{decom1} \vL_1 = \vL + \vR_1\qquad,\qquad \vL_2 = \vL + \vR_2\;, \ee
where $\vL \equiv \vx_\text{E}(t_1)-\vx_0$ is a vector directed from the
radio source to the geocenter having coordinates
$\vx_\text{E}=\vx_\text{E}(t_1)$, and $\vR_1 \equiv
\vx_1(t_1)-\vx_\text{E}(t_1)$, $\vR_2 \equiv \vx_2(t_1)-\vx_\text{E}(t_1)$
are the geocentric vectors of the first and second VLBI stations calculated
in the BCRS.

For an analytical treatment  one might employ a parallax
expansion of the quantities $\Delta t_0$ and $\vk_2$ with respect to the
powers of the small parameters $\epsilon_1\equiv R_1/L$ and $\epsilon_2\equiv
R_2/L$. These small parameters are of the order $\epsilon\simeq 2\times
10^{-2}$ for a radio source at the distance of the lunar orbit or smaller for
any other radio sources in the solar system.

For the parallax expansion of  $\vk_2$ we use the relation
\ba\la{legp1} \left(1-2\epsilon x + \epsilon^2\right)^{-1/2}
&=&\sum_{n=0}^{\infty}{P_n(x)\epsilon^n}  \,, \ea where $P_n(x)$ are the
usual Legendre polynomials.
For the parallax expansion of $\Delta t_0$ we employ the relation

\ba\la{gegenp1} \left(1-2\epsilon x +
\epsilon^2\right)^{1/2} &=&\sum_{n=0}^{\infty}{C_n(x)\epsilon^n} \, , \ea
where $C_n(x) \equiv C_n^{(-1/2)}$ are the Gegenbauer
polynomials with index $-1/2$:
(see Eq. 8.930 in Gradshteyn \& Ryzhik 1994):
\ba\la{cp1} C_0(x)&=&1\;,\nonumber\\\la{cp2} C_1(x)&=&-x\;,\\\la{cp3} nC_n(x)
&=& (2n-3) x C_{n-1} - (n-3)C_{n-2} \qquad (n\ge 2)\, . \nonumber\ea

We obtain the following expressions where terms of order less than $10\,$fs have been ignored:
\begin{align}\label{finiteeqone}
(\Delta t)_{\rm v} \simeq  &(\Delta t_0 + \Delta t_{\rm grav}) \Bigg\{ 1 -  \left(\boldsymbol{\sigma}_2\cdot{\dot\vx_2\over c}\right) \sum_{n=0}^{4}{P_n(\cos\theta_2)\left(\frac{R_2}{L}\right)^n} \nonumber \\
&+ \left(\boldsymbol{\sigma}_2\cdot{\dot\vx_2\over c}\right)^2 \left[1+2\cos\theta_2 \frac{R_2}{L}+(4\cos^2\theta_2-1)\frac{R_2^2}{L^2}\right]-\left(\boldsymbol{\sigma}_2\cdot{\dot\vx_2\over c}\right)^3\Bigg\} \nonumber \\
& - \frac 12 c^{-1} \boldsymbol{\sigma}_2 \cdot \ddot \vx_2 \Delta t_0^2 + \frac 12  c^{-1}L^{-1} \left\vert \boldsymbol{\sigma}_2 \times \dot \vx_2 \right\vert^2 \left(1+\cos\theta_2{ R_2\over L}\right)\Delta t_0^2 \, .
\end{align}
and
\begin{align}\label{finiteeqtwo}
c \Delta t_0 = &  L_2 - L_1 = \vert \vL+\vR_2 \vert - \vert \vL+\vR_1 \vert \nonumber \\
\approx  &-(\vk_\rE \cdot \vb) + \frac{1}{2L}\left(|\vn_2\times\vk_\rE|^2 {R_2}^2-|\vn_1\times\vk_\rE|^2 {R_1}^2\right) \nonumber \\
& +\sum_{n=3}^7{\frac{1}{L^{n-1}}\left[C_n(\cos\theta_2) {R_2} ^n - C_n(\cos\theta_1) {R_1} ^n\right]} \, ,
\end{align}
where
\begin{eqnarray}
\boldsymbol\sigma_2 &\equiv& \vk_\rE - \vn_2 \cdot (R_2/L) \\
\vn_i &\equiv& {\vR_i \over R_i} \\
\cos(\theta_i) &=& \vk_\rE \cdot \vn_i \, .
\end{eqnarray}
In (\ref{finiteeqtwo}) the parallax terms has been expanded up to the 7th order in Gegenbauer polynomials to achieve an accuracy of order $10\,$fs. For transferring the vacuum time-delay
in (\ref{finiteeqtwo}) from the BCRS to the GCRS, and including a tropospheric delay, the reader is referred to Section 2.1.

\section{Conclusions}
\begin{figure}[h!]
\begin{center}
\includegraphics[scale=0.5]{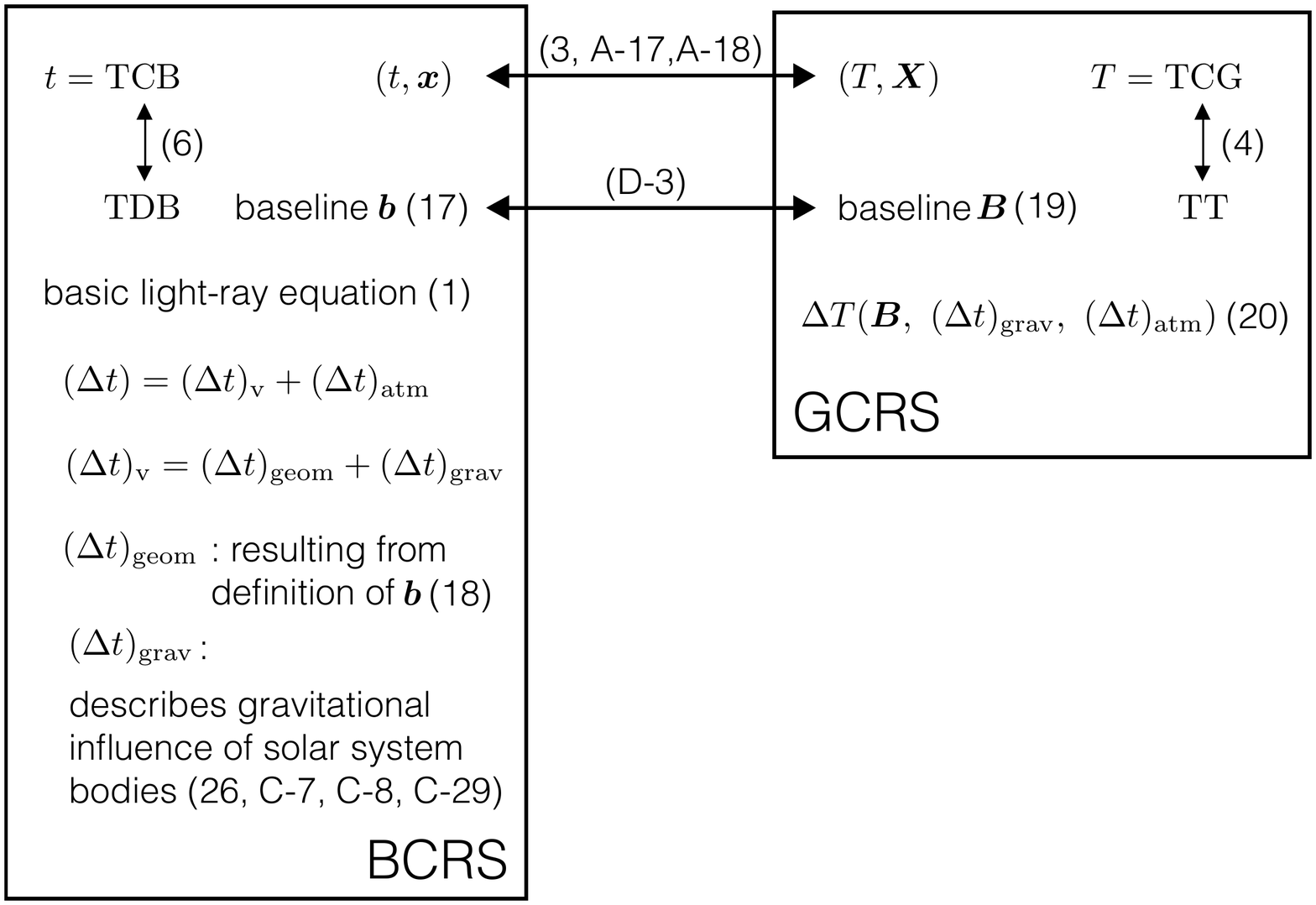}
\caption{Schematic diagram of the structure of the relativistic VLBI model for the group delay. The numbers refer to corresponding equations in the text
(definitions or relations).}\label{vlbidiagram}
\end{center}
\end{figure}

The purpose of this paper is a presentation of an advanced and fully-documented relativistic VLBI model for geodesy where earthbound baselines are considered. In contrast to the standard consensus model described in the IERS Conventions 2010,  our model is derived explicitly step by step from a well accepted formulation of relativistic celestial mechanics and astrometry.
A schematic diagram of the structure of our relativistic VLBI model for the group delay is presented in Fig.\ \ref{vlbidiagram}.

First, all terms from the consensus model are derived justifying this current standard model of VLBI data processing. However,
in various respects our model goes beyond the consensus model: terms related to the acceleration of the geocenter are included and arbitrary mass- and spin multipole moments are considered for the gravitating bodies in the problem of gravitational time delay (Shapiro delay) in general relativity. For the problem of radio sources located at finite distance a new parallax expansion is suggested here.
Thus, with the results from this paper realistic errors of the consensus model can be computed which is an essential theoretical addition to the IERS Conventions 2010.

For remote radio sources
a central result is the basic time delay equation (\ref{geom-transformed}) where the explicit form of the BCRS gravitational time delay is left open.  In principle, it can be derived from the results of Appendix C by means of formula (\ref{gravtimedelay}). The dominant terms resulting from the mass-monople, mass-quadrupole, spin-dipole and second post-Newtonian effects of some solar system body, that are already known from the literature, are written out explicitly.
Some orders of magnitude are presented in Table \ref{numbers} (Klioner 1991).

\begin{table} [h!]
\centering
  \caption{Maximal time-delays due to special effects from solar system bodies on VLBI observations with an earthbound baseline of $6000\,$km. } \label{numbers}
  \begin{tabular}{c|c|c|c|c|c}
   \\
  \hline\hline
   Body & 2pN & $J_2$ & $J_4$ & spin & motion\\
  \hline
  Sun & 307 ps  & 0.2 ps & $<$ 0.02 ps & 0.06 ps & 0.01 ps \\
  Jupiter & 1.5 ps & 21 ps & $<$ 1ps & 0.02 ps & 0.07 ps\\
  Saturn & 0.4 ps & 8 ps & -- & -- & 0.02 ps\\
  Uranus & 0.1 ps & 2 ps & -- & -- &-- \\
  Neptune & 0.3 ps & 0.7 ps & -- &-- &-- \\
  \hline\hline
  \end{tabular}
\end{table}

If the consensus model is extended to include effects from the mass-quadrupoles, spin-dipoles, 2PN effects and motion effects, then, it will be sufficient for most geodetic VLBI measurements also in the near future. We believe that the intrinsic accuracy of our model is of the order of $10\,$fs; further analyses will be made to check the orders of magnitude of all terms that have been neglected.

For some radio source at finite distance the main results are relations (\ref{finiteeqone}) and (\ref{finiteeqtwo}), where two different parallax expansions were employed. For more details on the problem of relativistic effects in the tropospheric delay the reader is referred to Kopeikin \& Han (2015).


\vfill\eject
\appendix

\section{ Appendix A: Theory of astronomical reference systems in brief}
\renewcommand{\theequation}{A-\arabic{equation}}
\setcounter{equation}{0}
A theory of relativistic reference system  has been formulated by Damour et al.\ (1991, DSX-I), improving and extending
earlier work by Brumberg \& Kopeikin (BK; 1989a,b). A standard reference is Soffel et al.\ (2003). For the VLBI model this theory provides precise definitions of the BCRS and the GCRS and the relations between them.
The importance of such two distinct reference systems results from relativity; even without gravity fields a geocentric baseline defined in the
barycentric system of coordinates would experience periodic relative variations  due to the  Lorentz contraction of order $10^{-8}$ which disappear completely in a suitably defined GCRS.
Note that also the basic time scales (TCG and TCB) for the BCRS and GCRS are different due to time dilation and gravitational redshifts.

The theory of astronomical reference systems that will be outlined below is formulated in the first post-Newtonian approximation of Einstein's theory of gravity.
The post-Newtonian approximation is a weak field slow motion approximation with small parameters $\epsilon_M = (GM/c^2 r)$ ($M$ being a mass of the system and $r$ the radial distance to $M$ in suitably chosen coordinates) and $\epsilon_v = (v/c)^2$ ($v$ being a translational or rotational coordinate velocity of a body or material element of the system). According to the Virial theorem one assumes that these two small parameters in the solar system have the same orders of magnitude, i.e., $\epsilon_M \sim \epsilon_v$ and one is using $c^{-1}$ as book-keeping parameter
for a post-Newtonian expansion of the metric tensor although it is not dimensionless.

Note, that what is called the first post-Newtonian approximation depends upon the problem of interest. If one talks about gravitational fields and celestial mechanical  problems of motion of massive bodies one neglects
terms of order $c^{-6}$ in the time-time component of the metric tensor, terms of order $c^{-5}$ in the time-space components and terms of order $c^{-4}$ in the space-space components.
If, however, we talk about the propagation of light rays we only consider $c^{-2}$ terms in the time-time component in the first PN approximation.

In what follows small letters refer to the BCRS, whereas capital letters
refer to the GCRS. We will  use  letters from the second part of the
greek alphabet like $\mu,\nu$ etc.\ for BCRS space-time indices and letters from the second part of the roman alphabet
like $i,j$ etc.\ for  BCRS spatial indices; we use greek indices like $\alpha,\beta$ etc.\ for GCRS space-time indices and roman indices
like
$a,b$ etc.\ for GCRS spatial indices.
The symbol $\cO(c^{-n})$ means that all terms of order $c^{-n}$ are neglected.

In Einstein's theory of gravity the gravitational field is described by a metric tensor that provides the geometry of spacetime.
In both systems, the BCRS and the GCRS, the metric tensor is written in the convenient form:
\begin{eqnarray}\label{basicm}
g_{00} = - e^{-2w/c^2} \, &;& \quad G_{00} = - e^{-2W/c^2} \, , \nonumber\\
g_{0i} = - {4 \over c^3} w^i \, &;& \quad G_{0a} = - {4 \over c^3} W^a \, , \\
g_{ij} = \gamma_{ij} e^{+2w/c^2} \, &;& \quad G_{ab} = \Gamma_{ab} e^{+2W/c^2} \, . \nonumber
\end{eqnarray}
Note, that in (\ref{basicm}) there is no approximation involved: the scalar potential $w$ replaces $g_{00}$, the vector potential $w^i$ replaces the time-space component of the metric tensor, $g_{0i}$, and the quantities $\gamma_{ij}$ replace the space components of the metric, $g_{ij}$.

Here, the $w$ and $W$ are the gravito-electric scalar potentials in the BCRS and GCRS respectively. They generalize
the usual Newtonian potential $U$; e.g., $w = U + \cO(c^{-2})$.

Einstein's field equations determine the components of the metric tensor (the gravitational field components) only up to four degrees of freedom. This gauge freedom corresponds to the free choice of the coordinate system.
We will use the harmonic gauge (e.g.\ Weinberg 1972) in every coordinate system. This has the consequence that (e.g.\ DSX-I)
\begin{equation}
\gamma_{ij} = \delta_{ij} + \cO(c^{-4}) \, ;  \ \Gamma_{ab} = \delta_{ab} + \cO(c^{-4}) \, ,
\end{equation}
so that the canonical form of the metric tensor in the first post-Newtonian approximation takes the form
\begin{eqnarray}\label{pnmetric}
g_{00} = - e^{-2w/c^2} \, &;& \quad G_{00} = - e^{-2W/c^2} \, , \nonumber \\
g_{0i} = - {4 \over c^3} w^i \, &;& \quad G_{0a} = - {4 \over c^3} W^a \, , \\
g_{ij} = \delta_{ij}\, e^{+2w/c^2} + \cO(c^{-4}) \, &;& \quad G_{ab} = \delta_{ab}\, e^{+2W/c^2} + \cO(c^{-4})  \, . \nonumber
\end{eqnarray}
In DSX-I the transformation $X^\alpha \rightarrow x^\mu$, i.e., from the GCRS to the BCRS is given, whereas the inverse transformation,
$x^\mu  \rightarrow X^\alpha$ is given in BK. The $X^\alpha \rightarrow x^\mu$ transformation is written in the form
\begin{equation} \label{gentrans}
x^\mu(T,X^a) = z^\mu(T) + e^\mu_a(T) X^a + \xi^\mu(T,X^a) \, ,
\end{equation}
where the functions $\xi^\mu$ are at least of order $\vert X^a\vert^2$. $z^i(T)$ describes the world-line of some central point associated with the Earth
that serves as origin of the spatial GCRS coordinates, i.e.,
\begin{equation}
z^i(T) = z^i_{\rm E}(T) \, ,
\end{equation}
parametrized with $T = \,$TCG.  For practical purposes it is convenient to choose  this central point as the post-Newtonian center of mass of the Earth. $z^0(T)$ is related with the $T \rightarrow
t$ transformation that will be discussed below.
We also introduce a vector quantity $e^\mu_0(T)$ by
\begin{equation}
e^\mu_0(T) \equiv  {1 \over c} {d z^\mu (T) \over dT} \, .
\end{equation}

Under general post-Newtonian assumptions (see DSX-I for more details)
one finds that
\begin{eqnarray}\label{xieq}
\xi^0(T,\vX) &=& \cO(c^{-3}) \, , \nonumber\\
\xi^i(T,\vX) &=& {1 \over c^2} e^i_a(T) \left[
{1 \over 2} A^a \vX^2 - X^a (\vA \cdot \vX) \right] + \cO(c^{-4}) \, ,
\end{eqnarray}
where $\vA$ is the coordinate acceleration of the geocenter projected into the local geocentric system.
The quantities $e^\mu_a(T)$ will be chosen as a tetrad field along the central world-line, orthonormal with respect to the external metric. Let us define
\begin{eqnarray}
\gbar_{00} (x) &\equiv& - e^{-2\wbar/c^2} \, , \nonumber \\
\gbar_{0i} (x) &\equiv& - {4 \over c^3} \wbar_i \, , \\
\gbar_{ij} (x) &\equiv& \delta_{ij} e^{+ 2\wbar/c^2} \, , \nonumber
\end{eqnarray}
where the external gravitational potentials, $\wbar$ and $\wbar_i$ describe the gravitational fields of all bodies other than the Earth:
\begin{equation}
\wbar = \sum_{B \not= {\rm E}} w_B \, \qquad  \wbar^i = \sum_{B \not= {\rm E}} w_B^i \, .
\end{equation}
The orthogonality condition for the quantities $e^\mu_a(T)$ then reads:
\begin{equation} \label{tetrads}
\gbar_{\mu\nu} e^\mu_\alpha e^\nu_\beta = \eta_{\alpha\beta} + \cO(6,5,4) \, .
\end{equation}
The order symbol $\cO(6,5,4)$ means that in $g_{00}$ terms of order $c^{-6}$, in  $g_{0i}$ terms of order $c^{-5}$ and in $g_{ij}$ of order $c^{-4}$ are neglected. From the tetrad conditions
(\ref{tetrads}) the following expressions can be derived (DSX I):
\begin{eqnarray}\label{tetex}
e^0_0 &=& 1 + {1 \over c^2} \left( {1 \over 2} \vv^2 + \wbar \right) \nonumber \\
&& \ \ \ + {1 \over c^4} \left[ {3 \over 8} \vv^4 + {1 \over 2} (\wbar)^2 + {5 \over 2} \wbar \vv^2 - 4 \wbar_i v^i\right] + \cO(c^{-6})  \nonumber \\
e^0_a &=& R^i_a \left[ {v^i \over c} \left[ 1 + {1 \over c^2} \left( {1 \over 2} \vv^2 + 3 \wbar \right)\right] - {4 \over c^3} \wbar_i \right] + \cO(c^{-5})
\\
e^i_a &=& \left( 1 - {1 \over c^2} \wbar \right) \left( \delta^{ij} + {1 \over 2 c^2} v^i v^j \right) R^i_a + \cO(c^{-4}) \, , \nonumber
\end{eqnarray}
in which the external potentials must be evaluated at the geocenter, i.e., the coordinate origin of the local system  at $X^a = 0$. $\vv$ is the barycentric coordinate velocity of the Earth,
\begin{equation}
v^i \equiv {dz^i \over dt}
\end{equation}
so that
\begin{equation}
e^i_0 = e^0_0 {v^i \over c} \, .
\end{equation}
In (\ref{tetex}) $R^i_a$ is a rotation matrix satisfying
\begin{equation}
R^i _a R^j _a = \delta^{ij} + \cO(c^{-4}) \, .
\end{equation}
The matrix decribes the orientation of the spatial GCRS coordnates, $X^a$, with respect to the barycentric ones, $x^i$. According to IAU 2000 resolution B1.3, we consider the GCRS to
be kinematically non-rotating with respect to the BCRS by writing
\begin{equation}
R^i_a(T) = \delta_{ia} \, .
\end{equation}

It is useful to invert the $X^\alpha \rightarrow x^\nu$ transformation and to write it as $x^\mu \rightarrow X^\alpha$. Note, that the solar system ephemerides are given in the BCRS with coordinates $x^\mu$.  To this end we have to keep in mind
that the general coordinate transformation (\ref{gentrans}) related the global coordinate $(ct,x^i)$ with the corresponding local
coordinates $(cT,X^a)$ of one and the same event $\cE$ in spacetime. Now, (\ref{gentrans}) involves functions $f(T)$ (e.g., $z^i(T), e^\mu_a(T)$ etc.)
defined at the central worldline. The $ T = {\rm const.}$ hypersurface through $\cE$ hits the central world-line $X^a = 0$  at a  point being different from the intersection
of this central worldline with the $t = {\rm const.}$ hypersurface. To derive the inverse transformation we have to set $f(t) = f(T_{\rm sim})$, where $T_{\rm sim}$ is the local coordinate time value of the intersection of the central world-line with the $t = $const.\ hypersurface. $T_{\rm sim}$ is related with the $T$ value of $\cE$ by
\begin{equation}
T^{\rm sim} = T + {\vv \cdot \vX \over c^2} + \cO(c^{-4}) \, .
\end{equation}
Details and an illustrating figure can be found in Appendix A of Damour et al.\ 1994.
With this the BK-transformation of spatial coordinates takes the form
\begin{equation}\label{trans-X}
X^a = \delta_{ai}\left\{ r^i + {1 \over c^2} \left[{1 \over 2} v^i (\vv \cdot \vr) + \wbar r^i + r^i (\va \cdot \vr) - {1 \over 2} a^i r^2 \right] \right\} + \cO(c^{-4}) \, ,
\end{equation}
where $\vr(t) \equiv \vx - \vz(t)$ and $\va$ is the coordinate acceleration of the Earth, $\va = d^2 \vz /dt^2$.
The transformation of time coordinates is more complicated. Basically the function $\xi^0$ is not fixed completely by the choice of harmonic
gauge in the BCSR and the GCRS. The IAU 2000 Resolution B.3  has chosen a special $T(t)$ transformation in accordance with (\ref{gentrans}):
\begin{eqnarray}\label{time-trans}
T &=& t - {1 \over c^2} \left[ A(t) + \vv \cdot \vr \right] \\
  && \ \ \ + {1 \over c^4} [B(t) + B^i(t) r^i + B^{ij}(t) r^i r^j + C(t,\vx)] + \cO(c^{-5}) \, , \nonumber
\end{eqnarray}
with
\begin{eqnarray*}
{d \over dt} A(t) &=& {1 \over 2} \vv^2 + \wbar  \, , \\
{d \over dt} B(t) &=& - {1 \over 8} \vv^4 - {3 \over 2} \vv^2 \wbar + 4 v^i \wbar^i + {1 \over 2} \wbar^2 \, , \\
B^i (t) &=& - {1 \over 2} \vv^2 v^i + 4 \wbar^i - 3 v^i \wbar \, \\
B^{ij} (t) &=& - v^i \delta_{aj} Q^a + 2 {\partial \over \partial x^j} \wbar^i - v^i {\partial \over \partial x^j} \wbar + {1 \over 2} \delta{ij}
\dot \wbar \, , \\
C(t,\vx) &=& - {1 \over 10} r^2 (\dot a^i r^i) \, .
\end{eqnarray*}
Here, the dot stands for the total derivative with respect to $t$, i.e.,
\begin{equation}
\dot \wbar \equiv \wbar_{,t} + v^i \wbar_{,i}
\end{equation}
and
\begin{equation}
Q^a(t) = \delta^a_i \left( {\partial \over \partial x^i} \wbar - a^i \right) \, .
\end{equation}
The quantity $Q^a$ describes a relative acceleration of the world-line of the Earth's center of mass with respect to the world-line of a freely falling, structure-less test particle.

\vfill\eject

%
%

\section{Appendix B: The metric tensor in the $N$-body system}
\renewcommand{\theequation}{B-\arabic{equation}}
\setcounter{equation}{0}
\subsection{The first post-Newtonian metric}

\subsubsection{The gravity field of a single body}
Let us consider the gravitational field a single body that we will describe in a single, global coordinate system $(t,\vx)$.
In harmonic gauge the field equations for the metric potentials, $w$ and $w^i$ take the form
\begin{eqnarray}
\block w &=& - 4 \pi G \sigma + \cO(c^{-4}) \, , \\
\Delta w^i &=& - 4 \pi G \sigma^i + \cO(c^{-2}) \, .
\end{eqnarray}
Here, $\block$ is the flat space d'Alembertian
\begin{equation}
\block \equiv - {1 \over c^2} {\partial^2 \over \partial t^2} + \Delta \, ,
\end{equation}
$\Delta$ is the Laplacian and
\begin{equation}
\sigma \equiv {T^{00} + T^{ss} \over c^2} \, , \qquad \sigma^i = {T^{0i} \over c} \, .
\end{equation}
The quantity $\sigma$ might be considered as the active gravitational mass-energy density and $\sigma^i$ as the active gravitational mass-energy current
that gives rise to gravito-magnetic effects described by the vector potential $w^i$ (the word 'active' refers to a field-generating quantity).
We can even combine the source- and the field  variables to form four-dimensional vectors
\begin{equation}
\sigma^\mu \equiv (\sigma, \sigma_i) \, , \qquad w^\mu \equiv (w,w_i)
\end{equation}
and write in obvious notation
\begin{equation}\label{FE}
\block w^\mu = - 4 \pi G \sigma^\mu + \cO(4,2) \, .
\end{equation}
We will consider an isolated system with
\begin{equation}
g_{\mu\nu} \rightarrow \eta_{\mu\nu} \qquad \vert \vx \vert \rightarrow \infty \, ,
\end{equation}
i.e.\ we consider our space-time manifold to be asymptotically flat. As is well known that under this condition the retarded and the advanced integrals
are solutions to the field equations (\ref{FE}):
\begin{equation}
w^\mu_{\rm ret/adv} (t,\vx) = G \int d^3 x'
{\sigma^\mu(t_{\rm ret/adv}; \vx') \over \vert \vx - \vx' \vert} \, ,
\end{equation}
with
\begin{equation}
t_{\rm ret/adv} \equiv t \mp \vert \vx - \vx' \vert/c \, .
\end{equation}
Another possible solution is
\begin{equation}
w^\mu_{\rm mixed} (t,\vx) = {1 \over 2} [w^\mu_{\rm ret} (t,\vx) + w^\mu_{\rm adv}(t,\vx)] \, .
\end{equation}
This mixed solution (the time symmetric solution) is in fact used in standard versions of the post-Newtonian formalism. The reason for that is the following:
if we expand $\sigma^\mu$ around the coordinate time $t$ we encounter a sequence of time derivative  terms and the term with the first derivative is related with time
irreversible processes such as the emission of gravity waves that do not occur in the first post-Newtonian approximation of Einstein's theory
of gravity. Therefore one chooses $w^\mu = w^\mu_{\rm mixed}$ with
\begin{equation}
w^\mu_{\rm mixed}(t,\vx) = G \int d^3x' {\sigma^\mu(t; \vx') \over \vert \vx - \vx' \vert} + {G \over 2 c^2} {\partial^2 \over \partial t^2}
\int d^3x' \, \sigma^\mu(t; \vx') \vert \vx - \vx' \vert \, .
\end{equation}
Next we shall discuss the gravitational field outside of some single body.

 It will be characterized with the help of Cartesian Symmetric and Trace Free (STF) tensors.
Here  special notations are used. Cartesian indices always run from $1$ to $3$ (or over $x,y,z$). Cartesian multi-indices, written with capital letters like $L$
will be used, meaning $L = i_1 i_2 \dots i_l$, i.e.\ $L$ stands for a whole set of $l$ different Cartesian indices $i_j$, each running from $1$ to $3$. E.g.\
\begin{equation*}
M_L = M_{i_1 \dots i_l}
\end{equation*}
has a total of $3^l$ components such as e.g., $M_{121}$ or $M_{233}$ for $l=3$. If one faces a Cartesian tensor that is symmetric in all indices and is free of traces such as
$M_{ssa}$ (summation of two equal dummy indices is always assumed), then it is called STF-tensor.
STF-tensors will be indicated with a caret on top or, equivalently, by a group of indices included in sharp parentheses:
\begin{equation}
\hat T_L  \equiv T_{<L>} \equiv T_{<i_1 \dots i_l>} \, .
\end{equation}
Mathematical algorithms how to get the STF part of an arbitrary tensor can be found in Thorne (1980) (see also DSX-I).

 To give an example, consider some arbitrary Cartesian tensor
$T_{ij}$. If we write a symmetrization of a group of certain indices with a round bracket as in
\begin{equation*}
T_{(ij)} = {1 \over 2} (T_{ij} + T_{ji})
\end{equation*}
then
\begin{equation*}
\hat T_{ij} = T_{(ij)} - {1 \over 3} \delta_{ij} \, T_{aa} \, .
\end{equation*}
An STF-tensor of especial importance is $\hat N_L$ with
\begin{equation}
n^L = n^{i_1 \dots i_l} = {x^{i_1} \dots x^{i_l} \over r^l} \, ,
\end{equation}
where $r^2 = x^s x^s = x^2 + y^2 + z^2$. The importance of $\hat n_L$ results from the fact that these quantities are equivalent to the usual scalar spherical harmonics
used in traditional expansions of the gravity field outside a body. In relativity, due to problems related with the Lorentz-transformation, expansions in terms of $\hat n_L$ are used rather than in terms of spherical harmonics. Since
\begin{equation}
\partial_L \left( \frac 1r \right) = (-1)^l (2l-1)!! {\nhat_L \over r^{l+1} } \,,
\end{equation}
in the exterior region of a body the gravitational potential $w$ in the Newtonian approximation can be written as
\begin{equation}
w(t,\vx) = G \sum_{l \ge 0} {(-1)^l \over l!} M_L \partial_L (r^{-1}) =  G \sum_{l \ge 0} {(2l-1)!! \over l!} M_L {\nhat_L \over r^{l+1}} \, ,
\end{equation}
where $M_L$ are the (Newtonian) Cartesian mass multipole-moments of the body.

A theorem due to Blanchet and Damour (1989) states: there is a special choice of harmonic coordinates
(called skeletonized harmonic coordinates) such that outside of some coordinate sphere that completely encompasses the matter distribution (body) the metric potentials
\begin{equation*}
w_\mu(t,\vx) = G \int d^3x' {\sigma_\mu(t_\pm; \vx') \over \vert \vx - \vx' \vert}
\end{equation*}
with
\begin{equation*}
t_\pm \equiv t_{\rm mixed} = {1 \over 2} (t_{\rm ret} + t_{\rm adv})
\end{equation*}
admit a convergent expansion of the form
\begin{eqnarray} \label{BD-expansion}
w(t,\vx) &=& G \sum_{l \ge 0} {(-1)^l \over l!} \partial_L [ r^{-1} M_L(t_\pm)] + \cO(c^{-4}) \nonumber \\
w_i(t,\vx) &=& - G\sum_{l \ge 1} {(-1)^l \over l!} \left[ \partial_{L-1}
\left( r^{-1} {d \over dt} M_{iL-1}(t_\pm) \right)\right. \nonumber \\&&  + \left. {l \over l+1} \epsilon_{ijk} \partial_{j L-1} (r^{-1} S_{kL-1}(t_\pm)) \right]
 + \cO(c^{-2}) \, .
\end{eqnarray}
Here, the mass-moments, $M_L$, and the spin-moments, $S_L$, are formally defined by
certain integrals of the body (Blanchet \& Damour, 1989; DSX-I). In the following these expressions, however, will not be needed. The external gravity potentials,
$w_\mu$ of a body are just determined by the set of multipole-moments.

\subsubsection{The metric tensor of $N$ moving bodies}

In the gravitational $N$-body problem we introduce a total of $N+1$ different coordinate systems: one global system of coordinates $(t,x^i)$ like the BCRS and a set of $N$ local
coordinates like the GCRS $(T_A, X^a_A)$ for each body $A$ of the system.

 Let us consider the metric tensor $G_{\alpha\beta}$ in the local system E (e.g., the Earth) defined by the two potentials $W_\alpha \equiv (W,W_a)$.
 In the gravitational $N$-body system these local potentials can be split into two parts
 \begin{eqnarray}
 \begin{split}
 W(T,\vX) &= W_{\rm self}(T,\vX)  + W_{\rm ext}(T,\vX) \\
 W^a(T,\vX) &= W^a_{\rm self}(T,\vX)  + W^a_{\rm ext}(T,\vX)
 \end{split}
 \end{eqnarray}
 with
 \begin{eqnarray}
 \begin{split}
 W_{\rm self} (T,\vX) &= G \int_\rE d^3X' {\Sigma(T,\vX') \over \vert \vX - \vX' \vert } + {G \over 2 c^2}
 {\partial^2 \over \partial T^2} \int_\rE d^3X' \Sigma(T,\vX') \vert \vX - \vX' \vert \, , \\
 W^a_{\rm self} (T, \vX) &= G \int_\rE d^3X' {\Sigma^a(T,\vX') \over \vert \vX - \vX' \vert} \, ,
 \end{split}
 \end{eqnarray}
 where the integrals extend over the support of body E only; $\Sigma$ is the gravitational mass-energy density in local E-coordinates, $\Sigma^a$ the corresponding gravitational mass-energy current. In DSX-I it was shown that the
 Blanchet-Damour theorem applies for the self-potentials when mass- and spin-multipole of body E, $M_L^\rE$ and $S_L^\rE$, are defined by corresponding intergrals
 taken over the support of body E only. This implies that in the local E-system the self-part of the metric outside the body E admits an expansion of the form
 (\ref{BD-expansion}) with BD-moments of body E.

In DSX-I it was also shown how to transform the self-parts of the local E-metric into the global system. Let
\begin{eqnarray}
\begin{split}
w_\rE(t,\vx) &= G \int_\rE d^3 x' {\sigma (t,{\vx'}) \over \vert \vx - \vx'\vert }
+ {G \over 2 c^2} {\partial^2 \over \partial t^2}
\int_\rE d^3 x' \sigma(t,\vx') \vert \vx - \vx' \vert \, , \\
w_\rE^i (t,\vx) &= G \int_\rE {\sigma^i(t,\vx') \over \vert \vx - \vx' \vert } \, ,
\end{split}
\end{eqnarray}
 then (DSX-I)
 \begin{eqnarray} \label{trans-self}
 \begin{split}
 w_\rE &= \left( 1 + {2 \vv^2 \over c^2} \right) W_{\rm self} + {4 \over c^2} v^a W^a_{\rm self} + \cO(c^{-4}) \\
 w^i_\rE &= \delta^i_a W^a_{\rm self} + v^i W_{\rm self} + \cO(c^{-2}) \, .
 \end{split}
 \end{eqnarray}
 This remarkable result says that the self-parts of the metric tensor simply transform with a post-Newtonian Lorentz-transformation.
 The metric of our system, composed of $N$ moving, extended, deformable, rotating bodies  is then obtained from
 \begin{equation}
 w^\mu(t,\vx) = \sum_A w^\mu_A(t,\vx) \, .
 \end{equation}
 E.g.\ for a system of mass monopoles, $M_L = 0$ for $l \ge 1$ and $S_L  =0$, we have
 \begin{equation*}
 W_A = GM_A/R_A \, ; \qquad W^a_A = 0
 \end{equation*}
 in the local $A$-system and from (53) we get
 \begin{eqnarray}
 \begin{split}
 w_A(t,\vx) &= \left( 1 + {2\vv_A^2 \over c^2}\right) {G M_A \over R_A} + \cO(c^{-4}) \\
 w_A^i(t,\vx) &= v_A^i {G M_A \over R_A} + \cO(c^{-2}) \, .
 \end{split}
 \end{eqnarray}
 To get the right hand side entirely in terms of  global coordinates one has to express the inverse local distance, $1/R_A$, in these. Using
 (26) one obtains:
 \begin{equation}
 {1 \over R_A} = {1 \over r_A}
 \left[ 1 - {\wbar(\vz_A) \over c^2} - {1 \over 2c^2} (\vv_A \cdot \vn_A)^2 - {1 \over 2 c^2} \va_A \cdot \vr_A \right] + \cO(c^{-4}) \, ,
 \end{equation}
 where $\vr_A(t) = \vx - \vz_A(t)$ and $\vn_A = \vr_A/\vert \vr_A \vert$.

 \subsection{The second post-Newtonian metric}

In the 2PN-approximation we consider only a single mass-monopole at rest. The corresponding metric for light-ray propagation, where we do not consider a $c^{-6}$ term in $g_{00}$, in harmonic coordinates reads up to terms of order $c^{-5}$
(e.g.\ Anderson \& DeCanio 1975; Fock 1964)
\begin{eqnarray}
g_{00} &=& - 1 + {2 w_\rA \over c^2}  - {2 w_\rA^2 \over c^4} \nonumber\\
g_{0i} &=& 0 \\
g_{ij} &=& \delta_{ij} \left( 1 + {2 w_\rA \over c^2} + {2 w_\rA^2 \over c^4}\right) + {1 \over c^4} q_{ij}^\rA \nonumber
\end{eqnarray}
with
\begin{equation*}
w_\rA = {G M_\rA \over r_\rA}
\end{equation*}
and
\begin{equation}
q_{ij}^\rA = {G^2 M_\rA^2 \over r_\rA^2} (n_\rA^i n_\rA^j - \delta_{ij}) \, .
\end{equation}
Here, at the 2PN level, the canonical form of the metric (\ref{basicm}) is not used.

\vfill\eject
%
%
\section{Appendix C: The gravitational time delay}
\renewcommand{\theequation}{C-\arabic{equation}}
\setcounter{equation}{0}
In Einstein's theory of gravity light-rays are geodesics of zero length (null-geodesics). Gravitational fields lead to a light-deflection and
a gravitational time delay. In VLBI it is the gravitational time delay that has to be considered and modelled at the necessary level of accuracy.

The gravitational time delay can be computed from the null condition, $ds^2 = 0$, along the light-ray. Writing $g_{\mu\nu}=\eta_{\mu\nu}+h_{\mu\nu}$ we get
\begin{displaymath}
dt^2  = {1 \over c^2} d\vx^2 + \left( h_{00} + 2 h_{0i} {dx^i \over dt} + {1 \over c^2} h_{ij} {dx^i \over dt} {dx^j \over dt} \right) \, dt^2
\end{displaymath}
or ($\vert h_{\mu\nu} \vert \ll 1$)
\begin{equation}
dt \simeq {\vert d \vx \vert \over c} + {\vert d \vx \vert \over 2c} (h_{\mu\nu} n^\mu n^\nu) \,,
\end{equation}
where we have inserted $\dot x^i = c n^i$ from the unperturbed light-ray equation, $\vx(t) = \vx_0 + \vn c(t - t_0)$ and $n^\mu \equiv (1,\vn)$.
For our metric, (\ref{pnmetric}), the Time Transfer Function (TTF), $\cT(t_0,\vx_0;\vx) \equiv t - t_0$ with $ds = \vert \vx \vert$ reads (e.g.\ Soffel \& Han 2015)
\begin{equation}\label{TTF}
 \cT(t_0,\vx_0; \vx) = \frac Rc + {1 \over 2c} \int_{s_0}^s (h_{\mu\nu} n^\mu n^\nu) ds =  \frac Rc + {2 \over c^3}
\int_{s_0}^{s} \left( w - \frac 2c \vw \cdot \vn \right) \, ds + \cO(c^{-4}) \,.
\end{equation}
The TTF allows the computation of $t$ if $t_0, \vx_0$ and $\vx$ are
given.

\subsection{A single gravitating body at rest}

We consider first a single body at rest at the origin of our coordinate system. Space-time outside of the body is assumed to be stationary ("time independent"; e.g.\ Soffel \& Frutos 2016).
Then the metric potentials outside the body take the form (\ref{BD-expansion}) with time independent mutipole moments $M_L$ and $S_L$.

We now use the a special parametrization of  the unperturbed light-ray (Kopei\-kin, 1997)
\begin{equation}
\vx_s = \vd + \vn \cdot s
\end{equation}
with $\vd \cdot \vn = 0$, i.e.\ $\vd = \vn \times (\vx \times \vn) = \vn \times (\vx_0 \times \vn)$ is the vector that points from the origin to the point of
closest approach of the unperturbed light-ray.
We then have $s = \vn \cdot \vx_s$.
Following Kopeikin (1997) we can now split the partial derivative with respect to  $x^i$ in the form
\begin{equation}
\partial_i = \partial^\perp_i + \partial^\parallel_i
\end{equation}
with
\begin{equation}
\partial^\perp_i \equiv {\partial \over \partial d_i} \, , \qquad
\partial^\parallel_i \equiv n^i {\partial \over \partial s} \, ,
\end{equation}
Then, (Kopeikin, 1997, equation (24)):
\begin{equation}
\partial_L = \sum_{p=0}^l {l! \over p! (l-p)!} n_P \partial^\perp_{L-P} \partial_s^p \, ,
\end{equation}
where $n_P = n_{i_1} \dots n_{i_p}$ and $\partial_s^p = \partial^p/\partial
s^p$. Inserting this into expression (\ref{TTF}) we get (the symbol $({\rm expr})_0$
stands for the expression taken at the initial point)
\begin{equation}\label{MMM}
\cT_{\rm M} = {2G \over c^3} \sum_{l=0}^\infty \sum_{p=0}^l {(-1)^l \over l!}
{l! \over p! (l-p)!} M_L n_P \partial^\perp_{L-P} \partial_s^p \ln {s + r
\over s_0 + r_0} - {\rm (expr)}_0
\end{equation}
for the time delay induced by the mass multipole moments $M_L$ and
\begin{equation}\label{SSS}
\cT_{\rm S} =  {4G \over c^4} \sum_{l = 1}^\infty \sum_{p=0}^l  {(-1)^l \over
l!} {l! \over p! (l-p)!}{l \over l+1} \epsilon_{ijk} S_{kL-1} n_P
\partial^\perp_{jL-P-1} \ln {s + r \over s_0 + r_0} - {\rm (expr)}_0
\end{equation}
for the time delay induced by the spin multipole moments $S_L$, since ($r_s = \vert \vx_s \vert$)
\begin{equation}
\int_{s_0}^{s} {ds \over r_s} =  \ln {s + r \over s_0 + r_0} \, .
\end{equation}
These results are in agreement with the ones found by Kopeikin (1997).
Let
\begin{equation}
\Phi(s,\vd) \equiv \ln(s + \sqrt{d^2 + s^2}) \,,
\end{equation}
then the first derivatives appearing in (\ref{MMM}) and (\ref{SSS}) read:
\begin{eqnarray}
\partial_s  \Phi &=&  \frac 1r \\
\partial^2_s \Phi &=& - {s \over r^3} \\
\partial^\perp_i \Phi &=& {d^i \over r(r+s)} \\
\partial^\perp\partial_s \Phi &=& - {d^i \over r^3} \\
\partial^{\perp}_{<ij>} \Phi &=& - {(s + 2r) \over (r+s)^2 r^3} d^i d^j - {n^i n^j \over r(r+s)} \, ,
\end{eqnarray}
where the last term results from the fact that (Kopeikin, 1997):
\begin{equation}
\partial^\perp_j d^i = \delta_{ij} - n^i n^j \, .
\end{equation}

\subsubsection{The mass-monopole moment}

Considering e.g.\ the mass-monopole term we have
\begin{equation*}
\cT_{{\rm M}, l=0} = 2 {G M \over c^3} \ln {s + r \over s_0 + r_0}
\end{equation*}
and since $s = \vn \cdot \vx_s $, we obtain
\begin{equation}\label{mass-monopole}
\cT_{{\rm M}, l=0} =  {2 G M\over c^3}
\ln \left( {\vn\cdot \vx +  r \over \vn \cdot \vx_0 +  r_0} \right) \,.
\end{equation}

\subsubsection{The mass-quadrupole}

\bigskip\noindent
For the mass-quadrupole  we get
\begin{equation} \label{quadrupolediff}
\cT_{{\rm M, l=2}} = {G \over c^3}  M_{ij} I_{ij}
\end{equation}
with
\begin{eqnarray}
I_{ij} &=& (n^i n^j \partial^2_s + 2 n^i \partial_s \partial_j^\perp + \partial^\perp_{ij}) \Phi \large\vert^s_0 \nonumber \\
&=& - n^i n^j \left( {s \over r^3} + {1 \over r(r+s)} \right) - {2 n^i d^j \over r^3} - {d^i d^j (s + 2r) \over (r+s)^2 r^3} \, .
\end{eqnarray}
Taking the integral expression for $\cT_{{\rm M, l=2}}$ one gets the same form  as in (\ref{quadrupolediff}) but with $I_{ij}$ being replaced by
\begin{eqnarray}
I'_{ij} &=&
3 \int_{s_0}^s {\frac{(d^i+n^i \tau)(d^j+n^j \tau)}{(d^2+s^2)^{5/2}}ds} \nonumber\\
&=&  \left(\frac{s^3}{r^3}\frac{n^i n^j}{d^2}-\frac{2n^i d^j}{r^3}+\frac{3s d^2+2s^3}{r^3}\frac{d^i d^j}{d^4}\right)\biggr|_{s_0}^{s} \, .
\end{eqnarray}
With some re-writing, using $d^2 = r^2  - s^2$, one finds that $I'_{ij} = I_{ij} + {\rm const.}$. Expression (\ref{quadrupolediff})  agrees with the one given by Klioner (Klioner 1991):
\begin{equation}
I_{ij} = n^i n^j \left( {\vn \cdot \vr \over r} \right)
(d^{-2} - r^{-2}) - {2 n^i d^j \over r^3} + {d^i d^j \over d^2} \left[ {\vn \cdot \vr \over r} (2 d^{-2} + r^{-2}) \right] \, .
\end{equation}

\bigskip

\subsubsection{The spin-dipole moment}

The contribution from the spin-dipole can be written in the form
\begin{equation}\label{spin-dipole}
\cT_{{\rm S}, l = 1}
 = - {2 G \over c^4} \epsilon_{ijk} n^i I_j S_k
\end{equation}
with
\begin{equation}
I_j = \partial_j \int {ds \over r_s} = (\partial^\perp_j + n^j \partial_s)  \ln {s+r \over s_0 + r_0}
\end{equation}
or, since the $n^j$-term does not contribute,
\begin{equation}
\vI =  - {\vd\, s \over d^2 r}  + {\vd\, s_0 \over d^2 r_0} \,.
\end{equation}

\subsection{The TTF for a body slowly moving with constant velocity}

Let us now consider the situation where the gravitating body (called A) moves
with a constant slow velocity $\vv_\rA$; we will neglect terms of order
$v_\rA^2$ in this section. Let us denote a canonical coordinate
system moving with body A, $X^\alpha =(cT,X^a)$ (see e.g.\ Damour et al.\
1991) and the corresponding metric potentials by $W$ and $W^a$. Under our
conditions the transformation from co-moving coordinates $X^\alpha$ to
$x^\mu$ is a linear Lorentz-transformation of the form ($\vbeta_\rA \equiv
\vv_\rA/c$):
\begin{equation}
x^\mu = z_\rA^\mu(T) + \Lambda^\mu_\alpha X^\alpha
\end{equation}
with $z_\rA^\mu \equiv (0,\vz_\rA(T))$ and $\Lambda^0_0 = 1, \Lambda^0_a = \beta_\rA^a, \Lambda^i_0 = \beta^i_\rA, \Lambda^i_a = \delta_{ia}$. A transformation of the co-moving
metric to the rest-system then yields (see also Damour et al.\ 1991)
\begin{eqnarray} \label{potinx}
w &=& W + {4 \over c} \vbeta_\rA \cdot  \vW \nonumber \\
w_i &=& W v_\rA^i + W_i \, .
\end{eqnarray}

In the following we will only consider a moving mass-monopole for which, in our approximation, $w = GM/r$ and $w^i = (GM/r) \cdot v^i$ so that the TTF
takes the form
\begin{eqnarray}
\cT(t_0,\vx_0;\vx) &=& \frac Rc + {2GM \over c^3} \int \left[  {(1 - 2 \vbeta_\rA \cdot \vn) \over r} \right]\, ds \nonumber \\
&=& \frac Rc + {2GM g_\beta \over c^3} \int { ds' \over r}  \,,
\end{eqnarray}
where $g_\beta \equiv 1 - \vbeta_\rA \cdot \vn$ and $s' = g_\beta s$.

We now
parametrize the unperturbed light-ray in the form
\begin{equation}
\vx_\tau = \vz_\rA + \vd_\beta + \vn_\beta  \tau \,,
\end{equation}
where $\vn_\beta \equiv \vg_\beta/g_\beta$, $\vg_\beta \equiv \vn - \vbeta_\rA$, and
 $\vd_\beta = \vn_\beta \times(\vr_\rA \times \vn_\beta)$ is perpendicular to $\vn_\beta$ so that $r_\rA(\tau) = \sqrt{ d_\beta^2 + \tau^2}$ and $\tau = \vr_\rA \cdot \vn_\beta$.
The TTF therefore for our mass-monopole in uniform motion takes the form
\begin{equation*}
\cT_{{\rm M}, l=0} = 2 {G M_\rA \over c^3} g_\beta \ln {r_\rA + \tau \over r_\rA^0 + \tau^0}
\end{equation*}
and since $\tau = \vn_\beta \cdot \vr_\rA = \vg_\beta \cdot \vr_\rA/g_\beta$, we obtain
\begin{equation}\label{moving-mass}
\cT_{{\rm M}, l=0} =  {2 G M_\rA\over c^3}g_\beta
\ln \left( {\vg_\beta \cdot \vr_\rA + g_\beta r_\rA \over \vg_\beta \cdot \vr_\rA^0 + g_\beta r_\rA^0} \right)
\end{equation}
in accordance with the results form the literature (e.g., Klioner \& Kopeikin 1992; Bertone et al.\ 2014).


\vfill\eject
\section{Appendix D: More technical details}
\renewcommand{\theequation}{D-\arabic{equation}}
\setcounter{equation}{0}

\subsection{The baseline equation}
To relate some BCRS baseline $\vb$ with the corresponding geocentric one, equations (\ref{BCRSbaseline}) and (\ref{GCRSbaseline}),
we now consider two events: $e_1$ is signal arrival time at antenna 1 with coordinates  $(T_1, \vX_1)$ in the GCRS and $(t_1,\vx_1)$ in the BCRS.
The second event $e_2$ will be the position of antenna $2$ at GCRS-time $T_1$, with coordinates $(T_1, \vX_2(T_1)$ in the GCRS and $(t_1^*, x_2(t_1^*)$ in the BCRS.
From (\ref{gentrans})
we get
\begin{eqnarray*}
x_1^i(t_1) &=& z_\rE^i(T_1) + e^i_a(T_1) X_1^a(T_1) + \xi^i(T_1,\vX_1) \\
x_2^i(t_1^*) &=& z_\rE^i(T_1) + e^i_a(T_1) X_2^a(T_1) + \xi^i(T_1,\vX_2) \, .
\end{eqnarray*}
From the time transformation we get $t_1^* = t_1 + \delta t^*$ with
\begin{equation}
\delta t^* = {1 \over c^2} (\vv_\rE \cdot \vB) + \cO(c^{-4})
\end{equation}
so that
\begin{equation}
x_2^i(t_1^*) = x_2^i(t_1) + {1 \over c^2} (\vv_\rE \cdot \vB) v_2^i +
\cO(c^{-4}) \, ,
\end{equation}
where $\vv_2 = \vv_\rE + \vV_2$ is the barycentric coordinate velocity of
antenna $2$. Finally, we get {a formula} for the baseline transformation in
the form
\begin{eqnarray}\label{baselineeq}
\vb = \vB &-& {1 \over c^2}
\left( {1 \over 2} (\vB \cdot \vv_\rE) \vv_\rE + (\vB \cdot \vv_\rE) \vV_2 + \wbar(\vz_\rE) \vB \right) \nonumber \\
&+& \Delta\vxi + \cO(c^{-4}) \, .
\end{eqnarray}

\subsection{Transformation of time intervals}

 Using the general time transformation we get
\begin{eqnarray}\label{tTt}
\Delta t &=& \Delta T + {1 \over c^2} \int_{t_1}^{t_2} dt' \left(
\wbar (\vz_\rE) + {1 \over 2} \vv_\rE^2 \right) \nonumber \\
&& + {1 \over c^2} \delta_{ia} v_\rE^i(T_2) X_2^a(T_2)
- {1 \over c^2} \delta_{ia} v_\rE^i(T_1) X_1^a + \cO(c^{-4}) \, .
\end{eqnarray}
To lowest order $\Delta T \simeq \Delta t \simeq - (\vB\cdot \vk)/c$ and we will approximate the integral in (\ref{tTt}) by the integrand times $\Delta t$,
using
\begin{eqnarray*}
v^i_\rE(T_2) &=& v_\rE^i - a_\rE^i \left( {\vB \cdot \vk \over c} \right) + \cO(c^{-2})\\
X_2^a(T_2) &=& X_2^a - V_2^a \left( {\vB \cdot \vk \over c} \right) + \cO(c^{-2})\, ,
\end{eqnarray*}
where $\vv_\rE \equiv \vv_\rE(T_1)$ etc. In this way we get the $\Delta t$-transformation equation in the form
\begin{eqnarray}
\label{deltateq}
\Delta T_{\rm geom} &=& \Delta t_{\rm geom} - {1 \over c^2} (\vB \cdot \vv_\rE) \nonumber \\
&& + {1 \over c^3} (\vB \cdot \vk)
\left[ \wbar(\vz_\rE) + {1 \over 2} \vv_\rE^2 + \va_\rE \cdot \vX_2 + \vv_\rE \cdot \vV_2 \right] + \cO(c^{-4}) \, ,
\end{eqnarray}
{and for the gravitational and atmospheric time-delay we have, to sufficient accuracy,
\begin{eqnarray}
\Delta T_{\rm grav} &=& \frac{\Delta t_{\rm grav}}{1+\frac{1}{c}\vk \cdot \vv_2} \\
\Delta T_{\rm atm} &=& \frac{\Delta t_{\rm atm}}{1+\frac{1}{c}\vk \cdot \vv_2}
\end{eqnarray}
}

\subsection{Independent derivation of the basic delay equation}

This basic time delay equation (\ref{geom-transformed}) can be derived directly without the introduction of some BCRS baseline $\vb$. One starts with expression (\ref{delt-geom}).
The left hand side is transformed with relation (\ref{deltateq}) and for the right hand side
we have
\begin{eqnarray*}
x^i_2(t_2) - x^i_1(t_1) &=& z_\rE^i(T_2) - z^i_\rE(T_1) \\
&& + e^i_a(T_2) X_2^a(T_2) - e^i_a(T_1) X_1^a(T_1) \\
&& + \Delta \xi^i + \cO_4 \\
&=& B^i + v_2^i (\Delta T) + {1 \over 2} a_2^i {(\vB \vk)^2 \over c^2} + \cO(c^{-3}) \, ,
\end{eqnarray*}
The right hand side of (\ref{delt-geom}) can therefore be written in the form
\begin{eqnarray*}
{\rm RHS} &=& - {1 \over c} \vB \cdot \vk - {1 \over  c} (\vk \cdot \vv_2) \Delta T
+ {1 \over c^3} (\vB \cdot \vk) \left[ \wbar(\vz_\rE) - {1 \over 2} (\vB \cdot \vk)(\vk \cdot \va_2) \right] \\
&& \ \ \
- {1 \over 2 c^3} (\vB \cdot \vv_\rE) (\vk \cdot \vv_\rE) - {1 \over c} \vk \cdot \Delta \vxi + \cO(c^{-4}) \, .
\end{eqnarray*}
The transformed equation can then be solved for $(\Delta T)$, considering  that there is one $(\Delta T)$-term on the left hand side and another one on the right hand
side. The result is again equation (\ref{geom-transformed}) above.

 \subsection{The problem of nearby radio sources}
 Let us start with the vacuum part of the time-delay, equation (\ref{vacvac}). We then make
 a Taylor expansion of $\vx_2(t_2)$ at $t_1$,
\begin{align}\label{talorx2}
\vx_2(t_2) = \vx_2(t_1) + \dot \vx_2(t_1) (t_2-t_1) + \frac12 \ddot \vx_2(t_1)(t_2-t_1)^2  \, ,
\end{align}
where the higher-order terms have been omitted. This is fully sufficient for the Earth-bounded VLBI measurements with a baseline $c(t_2-t_1)\sim 6000$ km because even for the most close case of a radio transmitter on the Moon, the third term in the right side of \eqref{talorx2} will  produce a time delay of the order of 3 fs (femtosecond),
which is two orders of magnitude smaller than the current precision of VLBI.
A substitution of  (\ref{talorx2}) into (\ref{vacvac}) then yields
\begin{align}
\Delta t = c^{-1}\Big[\big|\vL_2+\dot \vx_2(t_1) \Delta t + \frac12 \ddot \vx_2(t_1)\Delta t^2 \big|-\big|\vL_1\big|\Big] + \Delta t_{\rm grav} \label{geomtime2} \, ,
\end{align}
where for the sake of convenience we have suppressed the index ``geom'' in $\Delta t_\text{geom} $.
We now expand (\ref{geomtime2}) in a Taylor series with respect to $\Delta t$ keeping all terms up to the quadratic order. It gives us
\begin{align}
\Delta t = (\Delta t_0 + \Delta t_{\rm grav}) - c^{-1}(\vk_2\cdot \dot\vx_2)\Delta t - \frac12 c^{-1}(\vk_2\cdot\ddot \vx_2) \Delta t^2 + \frac12 c^{-1}L_2^{-1}|\vk_2 \times \dot \vx_2|^2 \Delta t^2\, , \label{geomtime3}
\end{align}
with $\vk_2 \equiv (\vx_0-\vx_2)/|\vx_0-\vx_2|= -\vL_2/L_2$ and  the arguments of $\vx_2, \dot \vx_2$, $\ddot \vx_2$ are taken at time $t_1$, and
$ \Delta t_0 = {L_2 - L_1 / c}\,$.
Eq. (\ref{geomtime3}) is a quadratic equation with a very small quadratic term so that it is more convenient to solve it by iteration. This yields
\begin{eqnarray} \label{deldeldel}
\Delta t & = &(\Delta t_0 + \Delta t_{\rm grav})\left( 1 + c^{-1} \vk_2 \cdot \dot\vx_2 \right)^{-1}
 - \frac 12 c^{-1} \vk_2 \cdot \ddot \vx_2 (\Delta t_0 + \Delta t_{\rm grav})^2  \nonumber\\
 &&+ \frac 12 c^{-1} L_2^{-1} \vert \vk_2 \times \dot \vx_2 \vert^2 (\Delta t_0 + \Delta t_{\rm grav})^2 \, .
\end{eqnarray}
For an analytical treatment one might employ a Taylor expansion of the denominator of the first term on the right hand side of (\ref{deldeldel}) which results in
equation (\ref{finitedisdelt}) above.

For the derivation of remaining results from Section 2.2 the following relations are useful.
 The Euclidean norm of vectors $\vL_1$, $\vL_2$ are \ba\la{eno1}
L_1&=& L\left[1-2(\vk_\rE\cdot\vn_1){R_1 \over L} +{R_1^2 \over
L^2}\right]^{1/2}\;,\\\la{eno2} L_2&=& L\left[1-2(\vk_\rE\cdot\vn_2){R_2
\over L} +{R_2^2 \over L^2}\right]^{1/2}\;, \ea and the corresponding unit
vectors are expressed by
\begin{eqnarray}
\vk_1 &=& \left(\vk_\rE - {R_1 \over L} \vn_1\right)\left[1-2(\vk_\rE\cdot\vn_1){R_1 \over L} +{R_1^2 \over L^2}\right]^{-1/2} \, , \label{N2N}\\
\vk_2 &=& \left(\vk_\rE - {R_2 \over L} \vn_2\right)\left[1-2(\vk_\rE\cdot\vn_2){R_2 \over L} +{R_2^2 \over L^2}\right]^{-1/2} \, , \label{N2N}
\end{eqnarray}
where $\vk_\rE \equiv -\vL/L$, $\vn_1 \equiv \vR_1/R_1$, $\vn_2 \equiv \vR_2/R_2$ are auxiliary unit vectors.

\comment{However, the coordinate system used in this section is BCRS, the geometrical and gravitational time delay gave here is in TCB scale. In practice, one must transfer the baseline from BCRS to GCRS by using Eq. (\ref{b-trans}), and time scale from TCB to TCG by Eq. (\ref{delt-trans}) then again to TT with Eq. (\ref{TT}). }

\vfill\eject

\vskip2truecm\noindent
{\bf Acknowledgement:} The work of S. Kopeikin has been supported by the grant No.14-27-00068 of the Russian Science Foundation (RSF). The work of W.-B. Han has been supported by the NSFC (No. 11273045) and Youth Innovation Promotion Association CAS. We would like to thank the anonymous referees for their suggestions to improve this manuscript.

\bigskip\noindent
\section{References}

\bn
Anderson J, DeCanio, T  (1975) Equations of hydrodynamics in general relativity in the slow motion approximation. Gen Rel Grav  6:  197--237

\bn
Bertone  S, Minazzoli  O, Crosta  M-T, Le Poncin-Lafitte, C, Vecchiato  A, Angonin, M
(2014) Time Transfer functions as a way to validate light propagation solutions for space astrometry.  Class Quant Grav  31:  015021

\bn
Blanchet  L, Damour  T (1989) Post-Newtonian generation of gravitational waves. Ann Inst H Poincar\'e  50: 377--408

\bn
Brumberg V  (1987)  Post-post Newtonian propagation of light in the Schwarzschild field.  Kin Fiz Neb  3:  8--13 (in russian)

\bn
Brumberg  V, Kopeikin, S (1989a)  Relativistic theory of celestial reference frames. In: Kovalevsky J, Muller I, Kolaczek B (eds.),
Reference Systems, Kluwer, Dordrecht: 115

\bn
Brumberg  V, Kopeikin, S  (1989b)  Relativistic reference systems and motion of test bodies in the vicinity of the Earth. Nuovo Cimento B 103: 63--98

\bn
Damour  T, Soffel M, Xu C  (1991) General-relativistic celestial mechanics. I. Method and definition of reference systems. Phys Rev  D  43: 3273--3307 (DSX-I)

\bn
Damour T, Soffel M, Xu C (1994) General-relativistic celestial mechanics. IV. Theory of satellite motion. Phys Rev D 49: 618--635

\bn
Eubanks  T (1991) Proceedings of the U.S.\ Naval Observatory Workshop
on Relativistic Models for Use in Space Geodesy, U.S.\ Naval Observatory,
Washington, D.C.,  June 1991

\bn
Finkelstein  A, Kreinovich  V, Pandey  S  (1983) Relativistic Reduction for Radiointerferometric Observables.  Ap Sp Sci 94: 233--247

\bn
Fock V A (1964) Theory of Space, Time and Gravitation, New York: Macmillan

\bn
Gradshteyn  I,  Ryzhik  I (1994) Table of integrals, series and products, Amsterdam: Academic Press

\bn
Hellings  R (1986)  Relativistic effects in astronomical timing
measurements, Astron J  91(3): 650--659. Erratum, ibid, 92(6), 1446

\bn IERS Conventions (2010) Petit G, Luzum B (eds.), IERS
Convention Centre, IERS Technical Note No. 36, Frankfurt am Main: Verlag des Bundesamtes f\"ur Kartographie und Geod\"asie

\bn
Jacobs  C et al. (2013) in: Proc.\ of Les Journ\'ees 2013, Syst\`emes de r\'ef\'erence spatio-temporels, Capitaine N (ed.), Paris, 51

\bn
Jacobsen R (2003) JUP230 orbit solutions

\bn
Klioner  S (1991) General relativistic model of VLBI observations. In
Proc.\ AGU Chapman Conf.\ on Geodetic VLBI: Monitoring Global Change, Carter
W (ed.), NOAA Rechnical Report NOS 137 NGS 49, American Geophysical
Union, Washington, D.C., 188

\bn
Klioner S (2003) Light propagation in the gravitational field of moving bodies by means of Lorentz transformation I. Mass monopoles moving with
constant velocities. Astron.\ Astrophys.\ 404: 783--787

\bn
Klioner  S, Kopeikin  S (1992) Microarcsecond astrometry in Space: Relativistic effects and reduction of observations. Astron J 104: 897--914

\bn
Klioner S, Soffel  M (2004): Refining the relativistic model for Gaia: cosmological effects in the BCRS. Proceedings of the Symposium "The Three-Dimensional Universe with Gaia", 4-7 October 2004, Observatoire de Paris-Meudon, France (ESA SP-576), 305

\bn
Kopeikin  S (1997)  Propagation of light in the stationary field of multipole gravitational lens. J Math Phys 38: 2587--2601

\bn
Kopeikin  S  (1990) Theory of relativity in observational radio
astronomy. Sov Astron  34: 5--9

\bn
Kopeikin  S,  Sch\"afer, G (1999) Lorentz covariant theory of light propagation in gravitational fields of arbitrary-moving bodies. Phys Rev D  60: 124002

\bn
Kopeikin  S, Mashhoon  B (2002) Gravitomagnetic effects in the propagation of electromagnetic waves in variable gravitational fields of arbitrary-moving and spinning bodies. Phys Rev D  65: 064025

\bn
Kopeikin  S, Han  W-B (2015) The Fresnel-Fizeau effect and the atmospheric time delay in geodetic VLBI. J Geod 89:  829--835

\bn
Meichsner J, Soffel M (2015) Effects on satellite orbits in the gravitational field
of an axisymmetric central body with a mass monopole
and arbitrary spin multipole moments, Cel Mech Dyn Ast 123: 1--12

\bn
Misner C, Thorne  K, Wheeler J A (1973)  Gravitation. Freeman and Company, New York.

\bn
Lambert  S  (2011) The first measurement of the galactic aberration by
the VLBI. Soci\'et\'e Francaise d'Astronomie et d'Astrophysique (SF2A) 2011,
Alecian G, Belkacem K, Samadi R,  Valls-Gabaud D (eds.)

\bn
Richter  G, Matzner  R (1983) Second-order contributions to
relativistic time-delay in the parametrized post-Newtonian formalism.
Phys Rev D  28:  3007--3012

\bn
Sekido M, Fukushima T (2006)  VLBI model for radio source at finite
distance. J Geod 86: 137--149

\bn
Soffel  M (1989) Relativity in Celestial Mechanics, Astrometry and
Geodesy. Springer, Berlin

\bn
Soffel M, M\"uller J, Wu X, Xu C (1991) Consistent relativistic VLBI model with picosecond accuracy. Astron J 101: 2306--2310

\bn
Soffel  M, Klioner  S, Petit  G, et al.  (2003) The IAU 2000
Resolutions  for  Astrometry, Celestial Mechanics, and Metrology in the  Relativistic  Framework: EXPLANATORY SUPPLEMENT. Astron J 126: 2687--2706

\bn
Soffel  M, Han  W-B  (2015)  The gravitational time delay in the field of a slowly moving body with arbitrary multipoles: Phys Lett A  379: 233--236

\bn
Soffel M, Frutos F (2016) On the usefulness of relativistic space-times for the description
of the Earth's gravitational field, J Geod: DOI 10.1007/s00190-016-0927-4

\bn
Thorne  K (1980) Multipole expansions of gravitational radiation.  Rev Mod Phys 52: 299--339

\bn
Titov  O, Lambert S, Gontier  A-M  (2011)  VLBI measurement of the
secular aberration drift. Astron Astrophys 529: A91

\bn
Ulrich R, Hawkins G (1980) The solar gravitational figure: $J_2$ and $J_4$. Final report. Report: NASA-CR-163881

\bn
Weinberg S  (1972)  Gravition and cosmology.  Wiley, New York

\bn
Zeller  G, Soffel  M, Ruder  H, Schneider  M (1986) Ver\"off.\ der
Bayr.\ Komm.\ f.d.\ Intern.\ Erdmessung, Astronomisch-Geod\"atische Arbeiten,
Heft Nr.\ 48:  218--236

\end{document}